\documentclass[journal,twocolumn]{IEEEtran}

\usepackage{color}
\definecolor{gray}{rgb}{0.5,0.5,0.5}

\usepackage{amsmath,amsthm}
\usepackage{amssymb,amsfonts}
\usepackage{graphicx}
\usepackage{url}
\usepackage{subfigure}

\newtheorem{theorem}{Theorem}

\newcommand{\argmax}[1]{\mathop{\arg\max}\limits_{#1}}
\newcommand{\vecm}[1]{{\rm vec}({#1})}
\newcommand{\diagm}[1]{{\rm diag}({#1})}

\newcommand{\etaknpost}[1]{{\eta}_{kn}(#1)}
\newcommand{\invm}[1]{(#1)^{-1}}
\newcommand{\logdet}[1]{\log\det(#1)}
\newcommand{\mean}[1]{\mathbb{E}\{#1\}}
\newcommand{\meanc}[2]{\mathbb{E}_{#1}\{#2\}}
\newcommand{\tdetalnpost}[1]{\tilde{\eta}_{ln}(#1)}

\newcommand{\tdetaknpost}[1]{\tilde{\eta}_{kn}(#1)}
\newcommand{\Lambdakn}[1]{\boldsymbol{\Lambda}_{kn}(#1)}
\newcommand{\tdLambdakn}[1]{\tilde{\boldsymbol{\Lambda}}_{kn}(#1)}
\newcommand{\tr}[1]{\mathrm{tr}(#1)}

\newcommand{\bdLambda}{\boldsymbol{\Lambda}}

\newcommand{\AknIterd}{\mathbf{A}_{kn}^{(d)}}

\newcommand{\BknIterd}{\mathbf{B}_{kn}^{(d)}}

\newcommand{\CknIterd}{\mathbf{C}_{kn}^{(d)}}
\newcommand{\ClnIterd}{\mathbf{C}_{ln}^{(d)}}
\newcommand{\Deltak}{\boldsymbol{\Delta}_{k}}

\newcommand{\DknIterd}{\mathbf{D}_{kn}^{(d)}}
\newcommand{\DnIterd}{\mathbf{D}_{n}^{(d)}}
\newcommand{\Ekn}{\mathbf{E}_{kn}}
\newcommand{\EknIterd}{\mathbf{E}_{kn}^{(d)}}

\newcommand{\FknIterd}{\mathbf{F}_{kn}^{(d)}}
\newcommand{\Gkn}{\mathbf{G}_{kn}}
\newcommand{\GknnIterd}{\mathbf{G}_{kn}^{(d)}}
\newcommand{\Gammakn}{\boldsymbol{\Gamma}_{kn}}

\newcommand{\Hkn}{\mathbf{H}_{kn}}
\newcommand{\Hknplus}{\mathbf{H}_{k(n+1)}}
\newcommand{\Hkmone}{\mathbf{H}_{k1}}

\newcommand{\IMt}{\mathbf{I}_{M_t}}

\newcommand{\IMk}{\mathbf{I}_{M_k}}
\newcommand{\Idk}{\mathbf{I}_{d_k}}
\newcommand{\Jkn}{\mathbf{S}_{kn}}
\newcommand{\JknIterd}{\mathbf{S}_{kn}^{(d)}}
\newcommand{\JlnIterd}{\mathbf{S}_{ln}^{(d)}}
\newcommand{\Kkmone}{\mathbf{K}_{k1}}
\newcommand{\Mk}{\mathbf{M}_k}
\newcommand{\Omegak}{\boldsymbol{\Omega}_k}
\newcommand{\olBknIterd}{\overline{\mathbf{B}}_{kn}^{(d)}}

\newcommand{\olCknIterd}{\overline{\mathbf{C}}_{kn}^{(d)}}
\newcommand{\olClnIterd}{\overline{\mathbf{C}}_{ln}^{(d)}}
\newcommand{\olDknIterd}{\overline{\mathbf{D}}_{kn}^{(d)}}
\newcommand{\olDnIterd}{\overline{\mathbf{D}}_{n}^{(d)}}
\newcommand{\olFknIterd}{\overline{\mathbf{F}}_{kn}^{(d)}}
\newcommand{\olcalRkn}{\overline{\mathcal{R}}_{kn}}

\newcommand{\Ponemn}{\mathbf{P}_{1n}}
\newcommand{\Ptwomn}{\mathbf{P}_{2n}}
\newcommand{\PKn}{\mathbf{P}_{Kn}}
\newcommand{\Pkn}{\mathbf{P}_{kn}}
\newcommand{\Pln}{\mathbf{P}_{ln}}
\newcommand{\PonemnIterd}{\mathbf{P}_{1n}^{(d)}}
\newcommand{\PtwomnIterd}{\mathbf{P}_{2n}^{(d)}}
\newcommand{\PKnIterd}{\mathbf{P}_{Kn}^{(d)}}
\newcommand{\PknIterd}{\mathbf{P}_{kn}^{(d)}}
\newcommand{\PlnIterd}{\mathbf{P}_{ln}^{(d)}}
\newcommand{\Phikn}{\boldsymbol{\Phi}_{kn}}
\newcommand{\Pikn}{\boldsymbol{\Pi}_{kn}}
\newcommand{\Piln}{\boldsymbol{\Pi}_{ln}}
\newcommand{\Rkn}{\mathbf{R}_{kn}}

\newcommand{\ckRknIterd}{\check{\mathbf{R}}_{kn}^{(d)}}
\newcommand{\Rln}{\mathbf{R}_{ln}}
\newcommand{\RknIterd}{\mathbf{R}_{kn}^{(d)}}

\newcommand{\Sigmakn}{\boldsymbol{\Sigma}_{kn}}
\newcommand{\Tkn}{\mathbf{T}_{kn}}
\newcommand{\Thetakn}{\boldsymbol{\Theta}_{kn}}
\newcommand{\Uk}{\mathbf{U}_{k}}
\newcommand{\Vk}{\mathbf{V}_k}
\newcommand{\Vkn}{\mathbf{V}_{kn}}
\newcommand{\VMt}{\mathbf{V}_{M_t}}
\newcommand{\Wkn}{\mathbf{W}_{kn}}
\newcommand{\Wknplus}{\mathbf{W}_{k(n+1)}}
\newcommand{\Xkmone}{\mathbf{X}_{k1}}
\newcommand{\Xlmone}{\mathbf{X}_{l1}}
\newcommand{\Xikn}{\boldsymbol{\Xi}_{kn}}
\newcommand{\Ymone}{\mathbf{Y}_{1}}
\newcommand{\Zmone}{\mathbf{Z}_{1}}

\newcommand{\bfC}{\mathbf{C}}
\newcommand{\cknIterd}{c_{kn}^{(d)}}
\newcommand{\cnIterd}{c_{n}^{(d)}}
\newcommand{\calGkn}{\mathcal{G}_{kn}}
\newcommand{\calRkn}{\mathcal{R}_{kn}}
\newcommand{\hatHkn}{\hat{\mathbf{H}}_{kn}}

\newcommand{\hatxkn}{\hat{\mathbf{x}}_{kn}}

\newcommand{\sz}{\sigma_z^2}
\newcommand{\sumnok}{\sum\limits_{l \neq k}^K}
\newcommand{\sumK}{\sum\limits_{k =1}^K}
\newcommand{\tdC}{\tilde{\mathbf{C}}}
\newcommand{\tdcalGkn}{\tilde{\mathcal{G}}_{kn}}
\newcommand{\tdSigmakn}{\tilde{\boldsymbol{\Sigma}}_{kn}}
\newcommand{\tdGammakn}{\tilde{\boldsymbol{\Gamma}}_{kn}}
\newcommand{\tdGammaln}{\tilde{\boldsymbol{\Gamma}}_{ln}}
\newcommand{\tdPhikn}{\tilde{\boldsymbol{\Phi}}_{kn}}

\newcommand{\tdHkn}{\tilde{\mathbf{H}}{}_{kn}}
\newcommand{\xln}{\mathbf{x}_{ln}}
\newcommand{\xkn}{\mathbf{x}_{kn}}
\newcommand{\ykn}{\mathbf{y}_{kn}}
\newcommand{\zkn}{\mathbf{z}_{kn}}
\newcommand{\zvarikn}{\mathbf{z}'_{kn}}
\usepackage{cite}

\ifCLASSINFOpdf
\else
\fi
\begin{document}
%
\title{Robust Transmission for Massive MIMO Downlink with Imperfect CSI}

\author{An-An Lu, \IEEEmembership{Member, IEEE,} Xiqi Gao, \IEEEmembership{Fellow, IEEE,} Wen Zhong
and  Chengshan Xiao, \IEEEmembership{Fellow, IEEE}, Xin Meng, \IEEEmembership{Member, IEEE}

\thanks{Manuscript received October 30, 2018; revised March 5, 2019; accepted April 11, 2019.
The work of A.-A. Lu,  X. Q. Gao and W. Zhong was supported by
National Natural Science Foundation of China under Grants 61801113, 61761136016, 61631018 and 61320106003,
National Science and Technology Major Project of China under Grant 2017ZX03001002-004,
the Natural Science Foundation of Jiangsu Province under Grant BK20180362,
and the Huawei Cooperation Project.
The work of C. Xiao was supported in part by US National Science Foundation under Grant ECCS-1827592.
The associate editor coordinating the review of this paper and approving it for publication was Prof. de Lamare, Rodrigo
\emph{(Corresponding author: Xiqi Gao).}}
\thanks{A.-A. Lu, X. Q. Gao and W. Zhong are with the National Mobile Communications Research Laboratory, Southeast University,
Nanjing, 210096 China, E-mail: aalu@seu.edu.cn, xqgao@seu.edu.cn, wzhong@seu.edu.cn.}
\thanks{C. Xiao is with the Department of Electrical and Computer Engineering, Lehigh University, Bethlehem, PA 18015. Email: xiaoc@lehigh.edu.}
\thanks{X. Meng is with the Wireless Network Research Department, Huawei Technologies Co., Ltd, Shanghai 201206, China,
Email:xmeng@ieee.org.}
\thanks{Copyright@2018 IEEE. Personal use of this material is permitted. Permission from IEEE must be obtained for all other uses, in any current or future media, including reprinting/republishing this material for advertising or promotional purposes, creating new collective works, for resale or redistribution to servers or lists, or reuse of any copyrighted component of this work in other works.}
 }


\maketitle

\begin{abstract}
In this paper, the design of robust linear precoders for the massive multi-input multi-output (MIMO) downlink with imperfect channel state information (CSI) is investigated.
The imperfect CSI for each UE obtained at the BS is modeled as statistical CSI under a jointly correlated channel model with both channel mean and channel variance information,  which includes the effects of channel
estimation error, channel aging and spatial correlation. The design objective is to maximize the expected weighted sum-rate.
By combining the minorize-maximize (MM) algorithm with the deterministic equivalent method,
an algorithm for robust linear precoder design is derived.
The proposed algorithm achieves a stationary point of  the expected weighted sum-rate maximization problem.
To reduce the computational complexity,  two low-complexity algorithms are then derived. One for the general case, and the other for the case when all the channel means are zeros.
For the later case, it is proved that the beam domain transmission is optimal, and thus the precoder design reduces to the power allocation optimization in the beam domain. Simulation results show that the proposed robust linear precoder designs apply to various mobile scenarios and achieve high spectral efficiency.
\end{abstract}


%
\IEEEpeerreviewmaketitle

\section{Introduction}
Massive multiple-input multiple-output (MIMO) \cite{marzetta2016fundamentals,lu2014overview} has been one of the key technologies of fifth generation (5G) wireless networks.
It provides huge potential capacity gains by employing a large number of antennas at a base station (BS) and supports multi-user MIMO (MU-MIMO) transmissions on the same time and frequency resource.
With massive antenna arrays at the BS, it is also possible to achieve high energy efficiency.
To alleviate the multi-user interference and improve the sum-rate performance, the precoders for all the UEs at the BS should be properly designed. In this paper, we focus on the precoder designs for massive MIMO downlinks.

Massive MIMO is an extension of conventional multi-user MIMO.
The precoder design for multi-user MIMO has seen significant attention in different forms over many years \cite{goldsmith2003capacity,peel2005vector,weingarten2006capacity,sadek2007leakage,gesbert2007single,christensen2008weighted,caire2010multiuser,shi2011iteratively,ngo2013energy,adhikary2013joint,sunbeam,wang2015asymptotic,park2015multi,liu2016two}. There exists two types of precoders: nonlinear precoders and linear precoders.
Although nonlinear precoders such as DPC \cite{weingarten2006capacity} can achieve optimal performance,
they are not suitable to massive MIMO due to the high complexity.
For practical consideration, we investigate linear precoder designs for massive MIMO in this paper.
The precoder designs are based on the available channel state information (CSI) at the BS.
When the BS has perfect CSI of all UEs, there exists the widely used regularized zero forcing (RZF) precoder \cite{peel2005vector},
the signal to leakage noise ratio (SLNR) precoder \cite{sadek2007leakage}, and the classic iterative weighted minimum mean square error (WMMSE) method \cite{christensen2008weighted,shi2011iteratively}.
Among the three precoders, the WMMSE precoder is designed according to the sum-rate maximization criterion.
Thus, the WMMSE precoder can achieve better performance than the RZF precoder and the SLNR precoder.

In massive MIMO systems, there exists many practical challenge, such as power amplifier nonlinearities \cite{qi2012power,zou2015impact}, transceiver I/Q imbalance \cite{hakkarainen2016analysis} and quantization errors \cite{jindal2006mimo}.
In this paper, we concern the impacts of the imperfect CSI.
In practical massive MIMO systems, perfect CSI at the BS are usually not available due to channel estimation error, channel aging, etc. Furthermore, different users usually have different moving speeds. Thus, we need to model the channel uncertainty first.
In the literature \cite{caire2010multiuser,pei2012masked,park2015multi,raeesi2018performance}, the channel uncertainty are often be constructed as a complex Gaussian random matrix with independent and identically distributed (i.i.d.), zero mean and unit variance entries.
In this paper, we propose to use a more realistic channel model for practical systems.
To describe the channel in practical systems more precisely, we consider the impacts of channel estimation, use the jointly correlated channel model to represent the spatial correlation, and
the widely used Gauss-Markov process \cite{mondal2006channel,member2007joint,mamat2018optimizing} to model the time evolution of the channel.
We consider a massive MIMO downlink where the \textit{a priori} CSI for each UE available at the BS before channel estimation is expressed as a jointly correlated channel model \cite{weichselberger2006stochastic} with only channel covariance information.
After channel estimation, we model the \textit{a posteriori} CSI for each UE at the BS as statistical CSI under a jointly correlated channel model with both channel mean and channel covariance information.
With the established model, we are able to describe the channel uncertainty more precisely.
On this basis, we investigate the precoder design for massive MIMO downlink transmission robust to the imperfect CSI at the BS.

If all the users are quasi-static, the established model reduces the case that the perfect CSI are known.
When all the users move fast, the BS only has channel covariance information.
In such case, there exists the beam division multiplex access (BDMA) transmission \cite{sunbeam} and the joint spatial division and multiplexing (JSDM) approach \cite{adhikary2013joint} that are designed by maximizing the sum-rate. In the BDMA transmission, the BS  serves multiple users via different beams simultaneously. In the JSDM approach, the users are partitioned into groups with approximately the same channel covariance eigenspace.
In conclusion, to maximize the sum-rate, the iterative WMMSE method can be used to design linear precoders for massive MIMO downlinks when the channel of all users are quasi-static. On the contrary, when all users are in medium or high mobility scenarios, the BDMA transmission or the JSDM approach can be used.
It is natural to ask whether there exist any unified linear precoding method which is robust against imperfect CSI and maximize the sum-rate for massive MIMO downlinks.
The goal of this paper is to answer this important question.

When the BS has imperfect CSI, the widely used RZF precoder can be extended to the robust RZF \cite{you2015pilot}. However, the performance of the robust RZF, especially at the high speed scenario, is still far from optimal.
In this paper, we combine the MM (minorize-maximize) algorithm \cite{hunter2004tutorial, sun2016mm}
and the deterministic equivalent method \cite{couillet2011random,alu2016free} to solve the problem of maximizing the expected weighted sum-rate
over the proposed channel model.
The MM algorithm is a widely used method to find the stationary points of complicated optimization problems. It substitutes a simple optimization problem for a difficult optimization problem.
Inspired by the weighted WMMSE method, we find a convex quadratic minorizing function of the objective function which can be used to apply the MM algorithm.
The optimal solution of the surrogate problem needs calculating the expected values of several random matrices with respect to the channel matrices based on the established \textit{a posteriori} channel model. However, the expected values of the random matrices are rather difficult to compute.
To avoid this issue, we use the deterministic equivalent method, which can be used to compute the approximations
of the matrix expectations needed.
Based on the obtained approximations, we propose an algorithm for robust linear precoder design. Furthermore, we derive two low-complexity algorithms by reducing the number of large dimensional matrix inversions and avoiding large dimensional matrix inversions, respectively.

The rest of this article is organized as follows.
The system model and problem formulation are presented in Section II. The robust linear precoder designs based on the deterministic equivalents are shown in Section III. Simulation results are provided in Section IV.
The conclusion is drawn in Section V. Proofs of Theorems are provided in Appendices.

{\it Notations}: Throughout this paper, uppercase and lowercase boldface letters are used for matrices and vectors, respectively. The superscripts $(\cdot)^*$, $(\cdot)^T$ and $(\cdot)^H$ denote the conjugate, transpose and conjugate transpose operations, respectively. ${\mathbb E}\{\cdot\}$ denotes the mathematical expectation operator. In some cases, where it is not clear, subscripts will be employed to emphasize the definition. The operators ${\rm{tr}}(\cdot)$
 and $\det(\cdot)$ represent the matrix trace and determinant, respectively.  The operator $\otimes$ denotes the kronecker product. The Hadamard product of two matrices $\mathbf{A}$ and $\mathbf{B}$ of the same dimensions is represented by $\mathbf{A} \odot \mathbf{B}$.
The $N \times N$ identity matrix is denoted by $\mathbf{I}_N$. The $(i,j)$-th entry of the matrix $\mathbf{A}$ is denoted by $[\mathbf{A}]_{ij}$.

\section{System Model and Problem Formulation}

\begin{figure}[b!]
\centering
\includegraphics[scale=0.5]{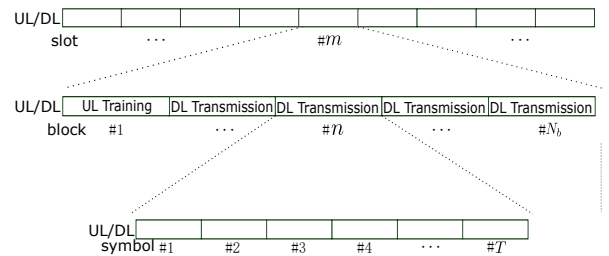}
\caption{Time slot structure.}
\label{fig:slot_structure}
\end{figure}

\subsection{System Model}
We consider a massive MIMO system with block flat fading channels, where the channel
coefficients remain constant for a coherence interval of $T$ symbol periods.
The system consists of one BS and $K$ UEs. The number of antennas at the BS is $M_t$. The $k$-th UE is equipped with $M_k$ antennas, and $\sum\nolimits_{k=1}^K M_k=M_r$.
We divide the time resources into slots and each time slot contains $N_b$ blocks.
In this paper, we focus on the case where the considered massive MIMO system operates in time division duplexing (TDD) mode. However, the results of this paper can be extended to the system operating in frequency division duplexing (FDD) mode easily.
For simplicity, we assume that there only exists the uplink training phase and the downlink transmission phase. At each slot, the uplink training sequences are sent once at the first block. The second block to the $N_b$-th block are used for downlink transmission.
The length of the uplink pilot sequences is $T$ symbols, \textit{i.e.}, the length of each block. Furthermore, the uplink training sequences assigned to different antennas are orthogonal to each other ($M_r \leq T$). For illustration purpose, we plot the time slot structure in Fig.~\ref{fig:slot_structure}.

We restrict our considerations to stationary channels and use the jointly correlated channel model to describe the spatial correlations of each channel. Specifically, the channel matrix $\Hkn$ from the BS to the $k$-th UE at the $n$th block of slot $m$ has the following structure \cite{weichselberger2006stochastic,gao:statistical}
    \begin{equation}
        \Hkn=\Uk(\Mk\odot\Wkn)\Vk^H
        \label{eq:channel_matrix_correlation_model_general}
    \end{equation}
where $\Uk$ and $\Vk$ are deterministic unitary matrices, $\Mk$ is an $M_k \times M_t$ deterministic matrix with nonnegative elements, and $\Wkn$ is a complex Gaussian random matrix with independent and identically distributed (i.i.d.), zero mean and unit variance entries. For brevity, we have omitted $m$ in the subscript. In this paper, we assume that uniform linear arrays (ULAs) are employed in the BS.
In such case, the covariance matrix $\mathbf{E}\{\Hkn^H\Hkn\}$ is a Toeplitz matrix under a wide sense stationary scattering environment.
When the number of the antennas at the BS grows large, the Toeplitz covariance matrix can be well approximated by a circulant matrix.
Thus, each $\Vk$ is closely approximated by a discrete Fourier transform (DFT) matrix.
Further, the channel model in \eqref{eq:channel_matrix_correlation_model_general} can be rewritten as
\begin{equation}
        \Hkn=\Uk(\Mk\odot\Wkn)\VMt^H
        \label{eq:channel_matrix_correlation_model}
\end{equation}
where $\VMt$ denotes the $M_t \times M_t$ DFT matrix. The channel model in $\eqref{eq:channel_matrix_correlation_model}$ can be seen as an \textit{a priori} model of the channels before channel estimation.
To model the time variation of the channel from block to block, we use the widely used first order Gauss-Markov process as in \cite{mondal2006channel,member2007joint,mamat2018optimizing}. Then, the channel matrix on the $n+1$-th block can be represented as
\begin{equation}
        \Hknplus = \alpha_k\Hkn + \sqrt{1-\alpha_k^2}\Uk(\Mk\odot\Wknplus)\VMt^H
        \label{eq:channel_matrix_temporal_correlation}
\end{equation}
where $\alpha_k$ is the temporal correlation coefficient which is related to the moving speed. An often used metric for $\alpha_k$ in the literature   \cite{member2007joint} is related to Jakes' autocorrelation model, \textit{i.e.}, $\alpha_k = J_0(2\pi v_k f_c  T/c)$,
where $J_0(\cdot)$ is the zero-th order Bessel function of the first kind,  $v_k$ is the moving speed of the $k$-th user, $f_c$ is the carrier frequency and $c$ is the speed of light.

We define the channel power matrices $\Omegak$ as $\Omegak=\Mk\odot\Mk$ and assume the BS knows $\Uk$ and $\Omegak$ through a channel sounding process.
Exploiting channel reciprocity, the channel state information of the downlink channels can be obtained from uplink training signals \cite{hoydis2013massive}.

Let $\Ymone^{\rm BS} \in \mathbb{C}^{M_t \times T}$ denote the received matrix at the BS on the first block of slot $m$. It can be written as
    \begin{IEEEeqnarray}{Cl}
        \Ymone^{\rm BS} &= \sumK \Hkmone^T\Xkmone^{\rm UE} + \Zmone^{\rm BS}
        \label{eq:received_pilot_model_in_matrix_form}
    \end{IEEEeqnarray}
where $\Xkmone^{\rm UE} \in \mathbb{C}^{M_k \times T}$ denotes the uplink training matrix sent by the $k$-th user on the first block of slot $m$, and $\Zmone^{\rm BS} \in \mathbb{C}^{M_t \times T}$ is a noise random matrix whose elements are i.i.d. complex Gaussian entries with zero mean and variance $\sigma_{\rm BS}^2$.

In the following, we model the \textit{a posteriori} CSI for each UE at the BS given $\Ymone^{\rm BS}$  as statistical CSI under a jointly correlated channel model with both channel mean and channel covariance information.
Vectorizing the received matrix $\Ymone^{\rm BS}$, we obtain
    \begin{IEEEeqnarray}{Cl}
        \!\!\!\!\vecm{\Ymone^{\rm BS}} &= \sumK \left((\Xkmone^{\rm UE} )^T \otimes \IMt \right)\vecm{\Hkmone^T} +  \vecm{\Zmone^{\rm BS}}.
        \label{eq:received_pilot_model_in_vector_form}
    \end{IEEEeqnarray}
Let $\Kkmone$ denote the covariance matrix of $\vecm{\Hkmone^T}$.
From \eqref{eq:channel_matrix_correlation_model}, we then obtain
\begin{IEEEeqnarray}{Cl}
        \Kkmone &=  \mean{\vecm{\Hkmone^T}\vecm{\Hkmone^T}^H} \nonumber \\
                &= (\Uk \otimes \VMt^*) \diagm{\vecm{\Omegak^T}}  (\Uk^H \otimes \VMt^T).
        \label{eq:covariance_matrix_of_vec_Hkmone}
    \end{IEEEeqnarray}
Let $\hatHkn^T$ denote the MMSE estimator of $\Hkn^T$ given $\Ymone^{\rm BS}$, we have that $\hatHkn$ is the conditional mean of $\Hkn^T$ given $\Ymone^{\rm BS}$, \textit{i.e.},
$\hatHkn^T = \mean{\Hkn^T|\Ymone^{\rm BS}}$.
Since the pilot sequences assigned to transmitted antennas are orthogonal, we obtain
$\Xkmone^{\rm UE}(\Xkmone^{\rm UE})^H = \IMk$ and $\Xlmone^{\rm UE}(\Xkmone^{\rm UE})^H = \mathbf{0}$ for $l \ne k$.
From these conditions, \eqref{eq:channel_matrix_correlation_model}, \eqref{eq:channel_matrix_temporal_correlation} and
\eqref{eq:received_pilot_model_in_vector_form}, we then obtain
    \begin{IEEEeqnarray}{Cl}
        \vecm{\hatHkn^T} &= \alpha_k^{n-1}\invm{\Kkmone +  \sigma_{\rm BS}^2\mathbf{I}}
        \nonumber \\
        &~~~~\Kkmone\left((\Xkmone^{\rm UE} )^* \otimes \IMt \right)\vecm{\Ymone^{\rm BS}}.
        \label{eq:mmse_estimator_of_hatHkn_in_vector_form_variant_1}
    \end{IEEEeqnarray}
Substituting \eqref{eq:covariance_matrix_of_vec_Hkmone} into \eqref{eq:mmse_estimator_of_hatHkn_in_vector_form_variant_1}, we obtain
    \begin{IEEEeqnarray}{Cl}
        &\vecm{\hatHkn^T}
        \nonumber \\
        &~~= \alpha_k^{n-1}(\Uk \otimes \VMt^*)\invm{ \diagm{\vecm{\Omegak^T}}
         +  \sigma_{\rm BS}^2\mathbf{I}}\nonumber\\
        &~~~~\diagm{\vecm{\Omegak^T}}\left(\Uk^H(\Xkmone^{\rm UE} )^* \otimes \VMt^T \right)\vecm{\Ymone^{\rm BS}}.
        \label{eq:mmse_estimator_of_hatHkn_in_vector_form_variant_2}
    \end{IEEEeqnarray}
Let $\Deltak$ denote the matrix whose entries are defined by
\begin{equation}
        [\Deltak]_{ij} =  \frac{[\Mk]_{ij}^2}{[\Mk]_{ij}^2+\sigma_{\rm BS}^2}.
        \label{eq:variance_of_Hkmone_given_Ymone}
\end{equation}
Then, \eqref{eq:mmse_estimator_of_hatHkn_in_vector_form_variant_2} can be re-expressed as
    \begin{IEEEeqnarray}{Cl}
        \vecm{\hatHkn^T} &= \alpha_k^{n-1}(\Uk \otimes \VMt^*)\diagm{\vecm{\Deltak^T}}
        \nonumber \\
        &~~~~\left(\Uk^H(\Xkmone^{\rm UE} )^* \otimes \VMt^T \right)\vecm{\Ymone^{\rm BS}}.
        \label{eq:mmse_estimator_of_hatHkn_in_vector_form_variant_3}
    \end{IEEEeqnarray}
From \eqref{eq:channel_matrix_correlation_model}, \eqref{eq:channel_matrix_temporal_correlation},
\eqref{eq:received_pilot_model_in_vector_form} and \eqref{eq:mmse_estimator_of_hatHkn_in_vector_form_variant_3}, we obtain the posterior distribution of the random vector $\vecm{\Hkn^T}$ is a multivariate Gaussian distribution, and its conditional covariance
matrix is given by
    \begin{IEEEeqnarray}{Cl}
       &\mean{(\vecm{\Hkn^T}-\vecm{\hatHkn^T})(\vecm{\Hkn^T}-\vecm{\hatHkn^T})^H |\Ymone^{\rm BS}}  \nonumber \\
       &~~=(\Uk \otimes \VMt^*)\diagm{\vecm{\Xikn^T \odot \Xikn^T}}(\Uk^H \otimes \VMt^T)
    \end{IEEEeqnarray}
where the square of the elements in $\Xikn \in \mathbb{C}^{M_k \times M_t}$ are computed by
\begin{equation}
        [\Xikn]_{ij}^2 =  [\Mk]_{ij}^2 - \alpha_k^{2(n-1)}\ \frac{[\Mk]_{ij}^4 }{[\Mk]_{ij}^2+\sigma_{\rm BS}^2}.
        \label{eq:variance_of_Hkn_given_Ymone}
\end{equation}
Finally, we obtain the \textit{a posteriori} model of $\Hkn$ given $\Ymone^{\rm BS}$ as
\begin{equation}
        \Hkn = \hatHkn + \Uk(\Xikn \odot \Wkn)\VMt^H
        \label{eq:posterori_model_of_Hkm}
\end{equation}
where $\hatHkn$ is obtained from \eqref{eq:mmse_estimator_of_hatHkn_in_vector_form_variant_3} as
\begin{equation}
    \hatHkn = \alpha_k^{n-1}\Uk(\Deltak \odot  \Uk^H(\Xkmone^{\rm UE} )^*(\Ymone^{\rm BS})^T\VMt)\VMt^H
\end{equation}
and $\Wkn$ is a complex Gaussian random matrix with i.i.d., zero mean and unit variance entries.
With \eqref{eq:posterori_model_of_Hkm}, the available imperfect CSI for each UE obtained at the BS is modeled
as statistical CSI under a jointly correlated channel model with both channel mean and channel variance information, which includes the effects of channel estimation error, channel aging and spatial correlation.
The channel model is obtained by assuming that $\Uk$, $\Omegak$, $\alpha_k$, $\sigma_{\rm BS}^2$ are known.
The \textit{a posteriori} model described in \eqref{eq:posterori_model_of_Hkm} is a generic model for  the available imperfect CSI obtained by the BS in the massive MIMO system under various mobile scenarios.
When $\alpha_k$ is very close to 1, it is suitable for the quasi-static scenario.
When $\alpha_k$ becomes very small, it is used to describe high speed scenario.
By setting the values of the $\alpha_k$s according to their moving speeds, we are able to describe the channel uncertainties in various typical channel conditions.
Based on this channel model, we investigate the precoder design robust to the imperfect CSI at the BS  in this work.

\subsection{Problem Formulation}
We now consider the downlink transmission for slot $m$.
Let $\xkn$ denote the $M_k \times 1$ transmitted vector to the $k$-th UE at the $n$-th block of slot $m$. The covariance matrix of $\xkn$ is
 the identity matrix $\Idk$.
The received signal $\ykn$ at the $k$-th UE for a single symbol interval at the $n$-th block of slot $m$ can be written as
    \begin{equation}
        \ykn = \Hkn\Pkn\xkn+\Hkn\sumnok\Pln\xln + \zkn
    \end{equation}
where $\Pkn$ is the $M_t \times d_k$ precoding matrix of the $k$-th UE, and $\zkn$ is a
complex Gaussian noise vector distributed as $\mathcal{CN}(0,\sz\IMk)$.

We assume that the UEs obtain the perfect CSI of their corresponding effective channel matrices $\Hkn\Pkn$ from the training signals as in the BDMA transmission \cite{sunbeam}.  The DL training phase is included in the DL data transmission and omitted in the slot structure for simplicity. At each UE, we treat the aggregate interference-plus-noise $\zvarikn = \Hkn\sum_{l\ne k}^K\Pln\xln + \zkn$ as Gaussian noise. Let $\Rkn$ denote the covariance matrix of $\zvarikn$, we have that
    \begin{equation}
        \Rkn = \sz\IMk+ \sumnok \meanc{}{\Hkn\Pln\Pln^H\Hkn^H}
        \label{eq:definition_of_covariance_matrices_of_effective_noise}
    \end{equation}
where the notation $\meanc{}{\cdot}$
denotes the expectation with respect to $\Hkn$ according to the long-term statistics of the channel matrices at the user end.
Owing to the channel reciprocity, the long-term channel statistics at the user end are the same as that of the BS, which have been provided in the \textit{a posteriori} model in \eqref{eq:posterori_model_of_Hkm}. Thus, the expectation $\meanc{}{\cdot}$ can be computed according to \eqref{eq:posterori_model_of_Hkm}.
We assume the covariance matrix $\Rkn$ is known at the $k$-th UE.
In such case,
the expected rate of the $k$-th user at the $n$-th block of slot $m$ is given by
    \begin{IEEEeqnarray}{Cl}
          \calRkn  &= \meanc{}{\logdet{\Rkn + \Hkn\Pkn\Pkn^H\Hkn^H  }} \nonumber \\
          &~~~~~~~~ -\logdet{\Rkn} \nonumber \\
                        &= \meanc{}{\logdet{\IMk + \Rkn^{-1}\Hkn\Pkn\Pkn^H\Hkn^H  }}
          \label{mutual_information}
    \end{IEEEeqnarray}
In this paper, we are interested in finding the precoding matrices $\Ponemn,\Ptwomn,\cdots,\PKn$ that maximize the expected weighted sum-rate. The optimization problem can be formulated as
    \begin{IEEEeqnarray}{Cl}
        &\Ponemn^{\diamond},\Ptwomn^{\diamond},\cdots,\PKn^{\diamond}
         \nonumber \\
        &~~=\argmax{\Ponemn,\cdots,\PKn}\sumK w_k\calRkn
        \nonumber \\
        &~~~~~~~~~~~~~~  {\rm s.t.} ~  \sumK \tr{\Pkn\Pkn^H} \leq P
        \label{eq:objective_optimization_problem}
    \end{IEEEeqnarray}
where $w_k$ are the weights to ensure fairness among users and $P$ denotes the total power budget.

\section{Robust Linear Precoder Design based on Deterministic Equivalents}

\subsection{MM Algorithm for Precoder Design}

In this subsection, we present the MM algorithm for precoder design.
The expected weighted sum-rate is a very complicated function of the precoding matrices, and thus
also very difficult to be optimized directly. In the following, we use the MM algorithm to find a stationary point of the optimization problem \eqref{eq:objective_optimization_problem}.

Let $f$ denote the objective function $\sum_{k=1}^K w_k\calRkn$ in the optimization problem
\eqref{eq:objective_optimization_problem}.  Let $\PonemnIterd,\PtwomnIterd, \cdots, \PKnIterd$ be a fixed family of the precoding matrices at the $d$-th iteration and let
    \begin{equation}
        g(\Ponemn,\Ptwomn, \cdots, \PKn|\PonemnIterd,\PtwomnIterd, \cdots, \PKnIterd) \nonumber
    \end{equation}
represent a real-valued continuous function of the precoders $\Ponemn,\Ptwomn, \cdots, \PKn$ whose form depends on the fixed precoding matrices $\PonemnIterd,\PtwomnIterd, \cdots, \PKnIterd$.
The function $g$ is said to minorize $f$ at $\PonemnIterd,\PtwomnIterd, \cdots, \PKnIterd$
provided that\cite{hunter2004tutorial}
\begin{IEEEeqnarray}{Cl}
        g(\Ponemn,\Ptwomn, \cdots, \PKn) & \leq  f(\Ponemn,\Ptwomn, \cdots, \PKn)
        \label{eq:minorize_condition_1}
\end{IEEEeqnarray}
where the equality holds at $\PonemnIterd,\PtwomnIterd, \cdots, \PKnIterd$.
When both the functions $g$ and $f$ are continuously differentiable with respect to the precoders, the conditions in \eqref{eq:minorize_condition_1} ensures
    \begin{IEEEeqnarray}{Cl}
        &\left.\frac{\partial g}{\partial \Pkn^*}\right|_{\Pkn=\PknIterd}
        =\left.\frac{\partial f}{\partial \Pkn^*}\right|_{\Pkn=\PknIterd}, k=1, \cdots, K.
        \label{eq:minorize_condition_3}
    \end{IEEEeqnarray}

The key of the MM algorithms for the considered problem is to obtain a surrogate function which minorize the objective function at any point.
When we find a good minorizing function, we will maximize it rather than the original function. Let $\Ponemn^{(d+1)},\Ptwomn^{(d+1)}, \cdots, \PKn^{(d+1)}$ denote the maximizer of
$g$ under the constraint.
From the condition \eqref{eq:minorize_condition_1}, we obtain
    \begin{IEEEeqnarray}{Cl}
        &f(\Ponemn^{(d+1)},\Ptwomn^{(d+1)}, \cdots, \PKn^{(d+1)}) \nonumber \\
        &~~~~~~\geq f(\PonemnIterd,\PtwomnIterd, \cdots, \PKnIterd).
        \label{eq:monotone_of_iterations}
    \end{IEEEeqnarray}
From \eqref{eq:minorize_condition_3} and \eqref{eq:monotone_of_iterations}, we observe that the sequence will converge to a local maximum
of the original function $f$. The proof of the convergence depends on the condition \eqref{eq:minorize_condition_1}  and has been provided in the literature \cite{vaida2005parameter,jacobson2007expanded}. Thus, we omit it here.

Let $\RknIterd$ and $\ckRknIterd$ be defined as
    \begin{IEEEeqnarray}{Cl}
        \RknIterd &= \sz\IMk+ \sumnok \meanc{}{\Hkn\PlnIterd(\PlnIterd)^H\Hkn^H}
        \label{eq:calculation_of_Rkn} \\
        \ckRknIterd &= \RknIterd +\Hkn\PknIterd(\PknIterd)^H\Hkn^H.
    \end{IEEEeqnarray}
With the previous definitions, we obtain the following theorem.
\vspace{-0.5em}
\begin{theorem}
\label{th:minorizing_function_1}
Let $g_1$ be a
function defined as
    \begin{IEEEeqnarray}{Cl}
        \!\!\!\!\!\!g_1&= \sumK w_k\cknIterd +  \sumK w_k\tr{\AknIterd\Pkn(\PknIterd)^H} \nonumber \\
        &~~+\sumK  w_k\tr{\AknIterd\PknIterd\Pkn^H}
       - \sumK \tr{\DknIterd\Pkn\Pkn^H}
        \label{eq:minorizing_function_definition_1}
    \end{IEEEeqnarray}
where  $\cknIterd$ is a constant provided in \eqref{eq:ckm_at_iterd}
and
    \begin{IEEEeqnarray}{Cl}
            \AknIterd&=\meanc{}{\Hkn^H\invm{\RknIterd}\Hkn}
            \label{eq:theorem_AkmIterd}
            \\
        \BknIterd&=\meanc{}{\Hkn^H(\invm{\RknIterd}-\invm{\ckRknIterd})\Hkn}
            \label{eq:theorem_BkmIterd}
            \\
        \CknIterd&= \meanc{}{\Hkn^H(\invm{\RknIterd}-\meanc{}{\invm{\ckRknIterd}})\Hkn} \label{eq:theorem_CkmIterd}
        \\
        \DknIterd &=  w_k\BknIterd + \sumnok w_l\ClnIterd
         \label{eq:theorem_DkmIterd}.
\end{IEEEeqnarray}
Then, it is a minorizing function of $f$ at $\PonemnIterd,\PtwomnIterd, \cdots, \PKnIterd$.
\end{theorem}
\vspace{-0.5em}
\begin{IEEEproof}
The proof is provided in Appendix \ref{sec:proof_of_minorizing_function_1}.
\end{IEEEproof}

Theorem \ref{th:minorizing_function_1} provides a minorizing function $g_1$ of the objective function.
Using the minorizing function $g_1$, we update
the precoding matrices sequence by
    \begin{IEEEeqnarray}{Cl}
        &\Ponemn^{(d+1)},\Ptwomn^{(d+1)}, \cdots, \PKn^{(d+1)}
        \nonumber \\
        &~~=\argmax{\Ponemn,\cdots,\PKn}g_1(\Ponemn,\Ptwomn, \cdots, \PKn)  \nonumber \\
        &~~~~~~~~~~~~~~{\rm s.t.} ~  \sumK \tr{\Pkn\Pkn^H} \leq P.
        \label{eq:update_through_surrogate_optimization_problem_1}
    \end{IEEEeqnarray}
The limit point of the sequence provided by \eqref{eq:update_through_surrogate_optimization_problem_1} is a stationary point of \eqref{eq:objective_optimization_problem}.
The optimization problem in \eqref{eq:update_through_surrogate_optimization_problem_1} is a concave quadratic optimization problem. Its optimal solution can be found by using the Lagrange
multiplier methods. We define the Lagrangian as
    \begin{IEEEeqnarray}{Cl}
        &\mathcal{L}(\mu, \Ponemn,\Ptwomn, \cdots, \PKn)\nonumber \\
        &~~~~  = - g_1+ \mu(\sumK \tr{\Pkn\Pkn^H} - P)
        \label{eq:lagrangian_of_surrogate_problem_1}
    \end{IEEEeqnarray}
where $\mu$ is the Lagrange multiplier.
From the first order optimal conditions of \eqref{eq:lagrangian_of_surrogate_problem_1}, we obtain
    \begin{IEEEeqnarray}{Cl}
              \Pkn^{(d+1)} &= (\DknIterd+\mu^\star\IMt)^{-1}w_k\AknIterd\PknIterd.
              \label{eq:optimal_solution_of_surrogate_optimization_1}
    \end{IEEEeqnarray}
Similar to that in \cite{shi2011iteratively}, the function $\sum_{k=1}^K{\rm tr}(\Pkn\Pkn^H)$ is a monotonically decreasing function of $\mu$.
Thus, if $\mu^\star=0$ and $\sum_{k=1}^K{\rm tr}(\Pkn^{(d+1)}(\Pkn^{(d+1)})^H)  \leq P$, we have obtained the optimal solution $\Pkn^{(d+1)}=(\DknIterd)^{-1}w_k\AknIterd\PknIterd$. Otherwise, we can obtain $\mu^\star$ by using a bisection method.

When the CSI is perfect known at the BS, the precoder obtained in \eqref{eq:optimal_solution_of_surrogate_optimization_1} reduces to the iterative WMMSE precoder.
Observing \eqref{eq:optimal_solution_of_surrogate_optimization_1},
we find the precoder $\PknIterd$ is first enhanced by $w_k\AknIterd$ and then filtered by $(\DknIterd+\mu^\star\IMt)^{-1}$.
From \eqref{eq:theorem_AkmIterd}, we find $\AknIterd$ can be seen as the expected weighted outer products of the channel column vectors of the $k$-th user, and $\DknIterd$ is dominated by the expected weighted outer products of the channel column vectors of the interference users.
Thus, $\AknIterd$ includes the information about the spatial directions that can be used to transmit the signal  for the $k$-th user, whereas $\DknIterd+\mu^\star\IMt$ includes the information about the spatial directions that will cause interference.
Furthermore, when $\hatHkn=\mathbf{0}$, $\AknIterd$ reduces to the weighted channel covariance matrix of
the $k$-th user, and $\DknIterd$ is dominated by the expected weighted channel covariance matrices of the interference users.
Using \eqref{eq:optimal_solution_of_surrogate_optimization_1}, we can obtain the precoders
that guarantee the gains of the signal and keep the interference small
at the same time.

To calculate the optimal solution in \eqref{eq:optimal_solution_of_surrogate_optimization_1}, we need to calculate $\AknIterd$, $\BknIterd$ and $\CknIterd$ using \eqref{eq:theorem_AkmIterd}, \eqref{eq:theorem_BkmIterd} and \eqref{eq:theorem_CkmIterd}.
Let
$\tdHkn$ denote $\Uk(\Xikn\odot \Wkn)\VMt^H$.
We define
\begin{IEEEeqnarray}{Cl}
\etaknpost{\tilde{\mathbf{C}}}&=\meanc{\tdHkn}{\tdHkn\tilde{\mathbf{C}}\tdHkn^H} \\
\tdetaknpost{\mathbf{C}}&=\meanc{\tdHkn}{\tdHkn^H\mathbf{C}\tdHkn}.
\end{IEEEeqnarray}
Then, we obtain
\begin{IEEEeqnarray}{Cl}
\meanc{}{\Hkn\tilde{\mathbf{C}}\Hkn^H}&=\hatHkn\tilde{\mathbf{C}}\hatHkn^H+\etaknpost{\tilde{\mathbf{C}}}
\label{eq:expectation_Hkm_c_HkmH} \\
\meanc{}{\Hkn^H\mathbf{C}\Hkn}&=\hatHkn^H\mathbf{C}\hatHkn + \tdetaknpost{\mathbf{C}}.
\label{eq:expectation_HkmH_c_Hkm}
\end{IEEEeqnarray}
From \eqref{eq:calculation_of_Rkn} and \eqref{eq:expectation_Hkm_c_HkmH}, we obtain
\begin{IEEEeqnarray}{Cl}
        \RknIterd &=\sz\IMk +\sumnok \hatHkn\PlnIterd(\PlnIterd)^H\hatHkn^H \nonumber \\
        &~~~~+\sumnok\etaknpost{\PlnIterd(\PlnIterd)^H}
        \label{eq:calculation_of_Rkn_variant}
\end{IEEEeqnarray}
From \eqref{eq:theorem_AkmIterd} and \eqref{eq:expectation_Hkm_c_HkmH}, we obtain $\AknIterd$ as
    \begin{equation}
        \AknIterd=\hatHkn^H\invm{\RknIterd}\hatHkn+\tdetaknpost{\invm{\RknIterd}}.
        \label{eq:calculation_of_Akn}
    \end{equation}
Equation \eqref{eq:theorem_BkmIterd} shows the first part of $\BknIterd$ can be obtained similarly as $\AknIterd$.       However, the second part of $\BknIterd$ is very complicated and it is rather difficult to obtain a closed-form expression.
Similarly, the computation of $\CknIterd$ also has no closed-form expression. In the next subsection, we will provide the approximations of $\BknIterd$ and $\CknIterd$ by using the deterministic equivalent method.


\subsection{Linear Precoder Design based on Deterministic Equivalents}
In this subsection, we provide a linear precoder design by using the deterministic equivalent method.
Observing \eqref{eq:theorem_BkmIterd} and \eqref{eq:theorem_CkmIterd}, we find that $\BknIterd$ and $\CknIterd$
are closely related to the derivatives of $\calRkn$ with respect to $\Pkn\Pkn^H$ and $\Pln\Pln^H$, $l \neq k$. Thus, to derive the deterministic equivalents of $\BknIterd$ and $\CknIterd$, we begin from the
deterministic equivalent of $\calRkn$. The channel model provided in \eqref{eq:posterori_model_of_Hkm}
is a jointly correlated channel model with a nonzero mean. For such model, the deterministic equivalent of $\calRkn$
has been provided in \cite{alu2016free} and \cite{wen2011on}.
Using the results from \cite{alu2016free}, we obtain the deterministic equivalent of $\calRkn$ as
\begin{IEEEeqnarray}{cl}
  \olcalRkn
   &= \logdet{\IMt + \Gammakn\Pkn\Pkn^H} + \logdet{\tdPhikn}
   \nonumber \\
   &~~~~- \tr{\etaknpost{\Pkn\calGkn\Pkn^H}\Rkn^{-1/2}\tdcalGkn\Rkn^{-1/2}}
  \label{eq:deterministic_equivalent_of_user_k_rate_1}
\end{IEEEeqnarray}
or
    \begin{IEEEeqnarray}{cl}
          \olcalRkn
            &= \logdet{\IMk + \tdGammakn\Rkn^{-1}} + \logdet{\Phikn}
           \nonumber \\
           &~~~~- \tr{\Pkn\calGkn\Pkn^H\tdetaknpost{\Rkn^{-1/2}\tdcalGkn\Rkn^{-1/2}}}
          \label{eq:deterministic_equivalent_of_user_k_rate_2}
    \end{IEEEeqnarray}
where $\Gammakn$ and $\tdGammakn$ are given by
\begin{IEEEeqnarray}{Cl}
\Gammakn&=\tdetaknpost{\Rkn^{-1/2}\tdcalGkn\Rkn^{-1/2}} \nonumber \\
        &~~~~+\hatHkn^H\Rkn^{-1/2}\tdPhikn^{-1} \Rkn^{-1/2}\hatHkn
\label{eq:calculation_of_Gammakn}
         \\
        \tdGammakn
            &=\etaknpost{\Pkn\calGkn\Pkn^H} + \hatHkn\Pkn\Phikn^{-1}\Pkn^H\hatHkn^H
\label{eq:calculation_of_tdGammakn}
\end{IEEEeqnarray}
and
$\Phikn$,  $\tdPhikn$,   $\calGkn$ and
$\tdcalGkn$ are obtained by the iterative equations
    \begin{IEEEeqnarray}{Cl}
        \Phikn &= \Idk +  \Pkn^H\tdetaknpost{\Rkn^{-1/2}\tdcalGkn\Rkn^{-1/2}}\Pkn
        \label{eq:calculation_of_Phikn}\\
        \tdPhikn &= \IMk + \Rkn^{-1/2}\etaknpost{\Pkn\calGkn\Pkn^H}\Rkn^{-1/2}
        \label{eq:calculation_of_tdPhikn}
        \\
        \calGkn &= \invm{\Idk + \Pkn^H\Gammakn\Pkn}
        \label{eq:calculation_of_calGkn}\\
        \tdcalGkn &= \invm{\IMk
        +  \Rkn^{-1/2}\tdGammakn\Rkn^{-1/2}}
        \label{eq:calculation_of_tdcalGkn}.
    \end{IEEEeqnarray}
From the two deterministic equivalents, we can obtain the derivatives of $\olcalRkn$
with respect to $\Pkn\Pkn^H$ and $\Pln\Pln^H$, $l \neq k$,
respectively.
With the obtained derivatives, we then obtain the deterministic equivalents of $\BknIterd$ and $\CknIterd$
in the following theorem.
\vspace{-0.5em}
\begin{theorem}
\label{th:deterministic_equivalent_of_mean_Bk_and_Ck}
The deterministic equivalents of $\BknIterd$ and $\CknIterd$ are
    \begin{IEEEeqnarray}{Cl}
        \olBknIterd &= \hatHkn^H\invm{\RknIterd}\hatHkn+\tdetaknpost{\invm{\RknIterd}}
          \nonumber \\
          &~~~~ -\invm{\IMt+\Gammakn\PknIterd(\PknIterd)^H}\Gammakn
        \label{eq:BkmIterd_deterministic_equivalent}
    \\
        \olCknIterd &=\hatHkn^H(\invm{\RknIterd}-\invm{\RknIterd+\tdGammakn})\hatHkn
        \nonumber \\
        &~~~~ +\tdetaknpost{\invm{\RknIterd}-\invm{\RknIterd+\tdGammakn}}.
        \label{eq:CkmIterd_deterministic_equivalent}
    \end{IEEEeqnarray}
\end{theorem}
\vspace{-0.5em}
\begin{IEEEproof}
The proof is provided in Appendix \ref{sec:proof_of_deterministic_equivalent_of_mean_Bk_and_Ck} .
\end{IEEEproof}
With the deterministic equivalents of $\BknIterd$ and $\CknIterd$ provided in Theorem \ref{th:deterministic_equivalent_of_mean_Bk_and_Ck},
the update step in \eqref{eq:optimal_solution_of_surrogate_optimization_1} using the minorizing function $g_1$ becomes
    \begin{IEEEeqnarray}{Cl}
              \Pkn^{(d+1)} &= (\olDknIterd+\mu^\star\IMt)^{-1}w_k\AknIterd\Pkn^{(d)}
              \label{eq:optimal_solution_of_surrogate_optimization_1_deterministic_equivalent}
    \\
              \olDknIterd &=  w_k\olBknIterd + \sumnok w_l\olClnIterd .
              \label{eq:deterministic_equivalent_of_Dkn}
\end{IEEEeqnarray}

From the computation of $\AknIterd$ in \eqref{eq:calculation_of_Akn}, the computation of
 $\olBknIterd$ and $\olCknIterd$ in Theorem \ref{th:deterministic_equivalent_of_mean_Bk_and_Ck} and the precoder in \eqref{eq:optimal_solution_of_surrogate_optimization_1_deterministic_equivalent}, it can be seen that the proposed method is directly derived from the \textit{a posteriori} channel mean and channel covariance information.
A relevant research, the stochastic weighted MMSE approach, can be find in \cite{razaviyayn2013stochastic}. It was extended from the iterative WMMSE method to maximize the ergodic sum rate for a MIMO interference channel.
It is an sample average approximation (SAA) method \cite{nemirovski2009robust, shapiro2014lectures}, which use a sample average problem to approximate the original optimization problem.

We now present an algorithm for the design of the robust linear precoder using the minorizing function $g_1$
with $\hatHkn$, $\Xikn$, $\Uk$ and $\sz$ as inputs.
\vspace{0.5em}
\hrule
\vspace{0.25em}
Algorithm 1: Linear precoder design using the minorizing function $g_1$
\vspace{0.25em}
\hrule
\begin{enumerate}[\IEEElabelindent=3em]
\item[Step 1:]
Set $d=0$. Randomly generate the precoders $\PonemnIterd,\PtwomnIterd,\cdots,\PKnIterd$ and normalize them to satisfy the power constraint.
\item[Step 2:]
Calculate $\RknIterd$ according to \eqref{eq:calculation_of_Rkn_variant}.
\item[Step 3:]
Calculate $\Gammakn$ and $\tdGammakn$  according to \eqref{eq:calculation_of_Gammakn} and
\eqref{eq:calculation_of_tdGammakn}.
\item[Step 4:]
Compute $\AknIterd$, $\olBknIterd$, $\olCknIterd$ and $\olDknIterd$ according to \eqref{eq:calculation_of_Akn},
\eqref{eq:BkmIterd_deterministic_equivalent}, \eqref{eq:CkmIterd_deterministic_equivalent} and
\eqref{eq:deterministic_equivalent_of_Dkn}.
\item[Step 5:] Update $ \Pkn^{(d+1)}$ by $\eqref{eq:optimal_solution_of_surrogate_optimization_1_deterministic_equivalent}$.
Set $d=d+1$.
\end{enumerate}
\hspace{1em}Repeat Step 2 through Step 5 until convergence or until a
pre-set target is reached.
\vspace{0.5em}
\hrule
\vspace{0.5em}
For very large $M_t$, the computational complexity of Algorithm 1 is dominated by the number of $M_t \times M_t$
matrix inversions. Observing Algorithm 1, we find there are an $M_t \times M_t$  inversion $\invm{\IMt+\Gammakn\PknIterd(\PknIterd)^H}$
in each computation of $\olBknIterd $ and an $M_t \times M_t$ inversion $(\olDknIterd+\mu^\star\IMt)^{-1}$
in each computation of $\Pkn^{(d+1)}$. Thus, there are total $2K$ $M_t \times M_t$ matrix inversions per iteration
and the computational complexity of Algorithm $1$ is of order ${\cal{O}}(K M_t^3)$ per iteration.

\subsection{Low-Complexity Linear Precoder Designs}
In this subsection, we present two low-complexity algorithms for linear precoder designs. The first algorithm is based on an alternative minorizing function modified from $g_1$. The second algorithm is designed for the case when $\hatHkn=\mathbf{0}$.

We begin with the first low-complexity algorithm. As shown in the previous subsection, the computational complexity of Algorithm 1 per iteration is dominated by $2K$ large dimensional matrix inversions.
The first $K$ large dimensional matrix inversions in Algorithm 1 can be avoid by rewriting them as
\begin{IEEEeqnarray}{Cl}
         \!\!\!\!\!\!&\invm{\IMt+\Gammakn\PknIterd(\PknIterd)^H}\Gammakn = \Gammakn \nonumber \\
          \!\!\!\!\!\!&~~~~ - \Gammakn\PknIterd\invm{\Idk+(\PknIterd)^H\Gammakn\PknIterd}(\PknIterd)^H\Gammakn
\end{IEEEeqnarray}
where the equality is due to the matrix inversion lemma.
The second $K$ large $M_t \times M_t$ matrix inversions $(\olDknIterd+\mu^\star\IMt)^{-1}$  can be reduced to one matrix inversion.
For this purpose, we provide the following theorem which presents an alternative minorizing function  modified from the minorizing function $g_1$.
\vspace{-0.5em}
\begin{theorem}
\label{th:minorizing_function_2}
Let $g_2$ be a
function defined as
    \begin{IEEEeqnarray}{Cl}
        g_2&=\cnIterd  +  \sumK \tr{(w_k\AknIterd+\FknIterd)\Pkn(\PknIterd)^H}
         \nonumber \\
        &~~~~+\sumK  \tr{(w_k\AknIterd+\FknIterd)\PknIterd\Pkn^H}
        \nonumber \\
        &~~~~~~~~~~~~~~- \sumK\tr{(\DknIterd+\FknIterd)\Pkn\Pkn^H}
        \label{eq:minorizing_function_definition_2}
    \end{IEEEeqnarray}
where $\FknIterd$ is any positive semidefinite matrix and $\cnIterd$ is a constant provided in \eqref{eq:constant_cn}.
Then, it is also a minorizing function of $f$ at $\PonemnIterd,\PtwomnIterd, \cdots, \PKnIterd$.
\end{theorem}
\vspace{-0.5em}
\begin{IEEEproof}
The proof is provided in Appendix \ref{sec:proof_of_minorizing_function_2}.
\end{IEEEproof}

The minorizing function $g_2$ can be used to reduce the complexity of the solutions of the surrogate optimization problem. Let $\FknIterd$ be defined as
   \begin{IEEEeqnarray}{Cl}
        \FknIterd &=  w_k\CknIterd -  w_k\BknIterd
    \end{IEEEeqnarray}
which is obviously a positive definite matrix, then we have
    \begin{IEEEeqnarray}{Cl}
        \DknIterd + \FknIterd = \sumK w_k\CknIterd
    \end{IEEEeqnarray}
which is the same for all $k$. For brevity, we define $\DnIterd = \sum_{k=1}^K w_k\CknIterd$.
From \eqref{eq:BkmIterd_deterministic_equivalent} and \eqref{eq:CkmIterd_deterministic_equivalent}, we obtain the deterministic equivalents of $\DnIterd$ and $\FknIterd$ as
   \begin{IEEEeqnarray}{Cl}
        \olDnIterd &=  \sumK w_k\olCknIterd
        \label{eq:DmIterd_deterministic_equivalent} \\
        \olFknIterd &=  w_k\olCknIterd -  w_k\olBknIterd \label{eq:FkmIterd_deterministic_equivalent}
    \end{IEEEeqnarray}
where the computation of $\olBknIterd$ becomes
    \begin{IEEEeqnarray}{Cl}
        \!\!\!\!\!\!\!\!\olBknIterd &= \hatHkn^H\invm{\RknIterd}\hatHkn+\tdetaknpost{\invm{\RknIterd}} -\Gammakn
    \nonumber \\
    &~ \!+ \!\Gammakn\PknIterd\invm{\Idk\!+\!(\PknIterd)^H\Gammakn\PknIterd}(\PknIterd)^H\Gammakn.
    \label{eq:calculation_of_Bkn_in_algorithm_2}
    \end{IEEEeqnarray}

The process of using the minorizing function $g_2$ to obtain a stationary point is similar to that of using $g_1$. For brevity, we omit the details and give the solution directly as
    \begin{IEEEeqnarray}{Cl}
              \Pkn^{(d+1)} &= (\olDnIterd+\mu^\star\IMt)^{-1}(w_k\AknIterd +\FknIterd)\PknIterd.
              \label{eq:optimal_solution_of_surrogate_optimization_2}
    \end{IEEEeqnarray}
The above equation is still similar to that of \eqref{eq:optimal_solution_of_surrogate_optimization_1}.
Thus, it also can  achieve a good performance.

We now present an algorithm for the design of the robust linear precoder using the minorizing function $g_2$
with $\hatHkn$, $\Xikn$, $\Uk$ and $\sz$ as inputs.
\vspace{0.5em}
\hrule
\vspace{0.25em}
Algorithm 2: Linear precoder design using the minorizing function $g_2$
\vspace{0.25em}
\hrule
\begin{enumerate}[\IEEElabelindent=3em]
\item[Step 1:]
Set $d=0$. Randomly generate the precoders $\PonemnIterd,\PtwomnIterd,\cdots,\PKnIterd$ and normalize them to satisfy the power constraint.
\item[Step 2:]
Calculate $\RknIterd$ according to \eqref{eq:calculation_of_Rkn_variant}.
\item[Step 3:]
Calculate $\Gammakn$ and $\tdGammakn$  according to \eqref{eq:calculation_of_Gammakn} and
\eqref{eq:calculation_of_tdGammakn}.
\item[Step 4:]
Compute $\AknIterd$, $\olBknIterd$, $\olDnIterd$ and $\olFknIterd$  according to
\eqref{eq:calculation_of_Akn},
\eqref{eq:calculation_of_Bkn_in_algorithm_2},  \eqref{eq:DmIterd_deterministic_equivalent} and \eqref{eq:FkmIterd_deterministic_equivalent}.
\item[Step 5:] Update $ \Pkn^{(d+1)}$ by $\eqref{eq:optimal_solution_of_surrogate_optimization_2}$.
Set $d=d+1$.
\end{enumerate}
\hspace{1em}Repeat Step 2 through Step 5 until convergence or until a
pre-set target is reached.
\vspace{0.5em}
\hrule
\vspace{0.5em}

For Algorithm 2, there only need one $M_t \times M_t$ matrix inversion per iteration and its computational complexity is of order ${\cal{O}}(\frac{1}{2}M_t^3)$ per iteration. Thus, the complexity of
Algorithm 2 is reduced compared with that of Algorithm 1 for very large $M_t$. To further reduce the computational complexity, the truncated conjugate gradient (CG) method can be used to solve \eqref{eq:optimal_solution_of_surrogate_optimization_2}.

In the following, we introduce another low-complexity algorithm for a special case.
When $\alpha_k^{n-1}$ is small, the elements in the \textit{a posteriori} channel mean $\hatHkn$ are also small.
In such case, knowing $\hatHkn$ can not bring much performance gain.
To reduce complexity, we can assume $\alpha_k^{n-1}=0$ and only use the \textit{a priori} channel information in \eqref{eq:channel_matrix_correlation_model}, which is equivalent to $\hatHkn=\mathbf{0}$ and $\Xikn=\Mk$. Then,
we obtain
\begin{IEEEeqnarray}{Cl}
    \AknIterd&=\tdetaknpost{\invm{\RknIterd}}  \\
    \olBknIterd &= \tdetaknpost{\invm{\RknIterd}} \nonumber \\
    &~~~~-\invm{\IMt+\Gammakn\PknIterd(\PknIterd)^H}\Gammakn \\
    \olCknIterd &= \tdetaknpost{\invm{\RknIterd}}  - \Gammakn
\end{IEEEeqnarray}
where the last equation is obtained by using the formula $\Gammakn=\tdetaknpost{\invm{\RknIterd+\tdGammakn}}$.
Furthermore, the computations of $\Gammakn$ and $\tdGammakn$ become
\begin{IEEEeqnarray}{Cl}
        \Gammakn
            &=\tdetaknpost{\Rkn^{-1/2}\tdcalGkn\Rkn^{-1/2}}
            \label{eq:Gammakm_zero_mean_case}
         \\
        \tdGammakn
            &=\etaknpost{\Pkn\calGkn\Pkn^H}
            \label{eq:tdGammakm_zero_mean_case}
    \end{IEEEeqnarray}
and $\tdPhikn, \Phikn, \calGkn$ and
$\tdcalGkn$ are now obtained by the iterative equations from \eqref{eq:calculation_of_Phikn}
to \eqref{eq:calculation_of_tdcalGkn} by setting $\hatHkn=\mathbf{0}$.
Let $\Lambdakn{\tdC}$ and $\tdLambdakn{\bfC}$ be two diagonal matrix valued functions defined as
\begin{IEEEeqnarray}{Cl}
[\Lambdakn{\tdC}]_{ii}&=\sum\limits_{j=1}^{M_t}[\Omegak]_{ij} [\VMt^H\tdC\VMt]_{jj}
\label{eq:eta_function_component} \\
\left[\tdLambdakn{\bfC}\right]_{ii}%
&=\sum\limits_{j=1}^{M_k}[\Omegak]_{ji} [\Uk^H\bfC\Uk]_{jj}.
\label{eq:tdeta_function_component}
\end{IEEEeqnarray}
Then, we obtain
\begin{IEEEeqnarray}{Cl}
\etaknpost{\tilde{\mathbf{C}}}&=\Uk\Lambdakn{\tdC}\Uk^H
\label{eq:etakmpri_zero_mean}
\\
\tdetaknpost{\mathbf{C}}&= \VMt\tdLambdakn{\bfC}\VMt^H.
\label{eq:tdetakmpri_zero_mean}
\end{IEEEeqnarray}

In such case, we observe from Algorithms 1 and 2 that once the left singular vector matrix of
$\Pkn^{(d)}$ is $\VMt$ right multiplying a permutation matrix, it will remain the same forever. Thus, we obtain that $\VMt$
right multiplying a permutation matrix must be the left singular vector matrix for $\Pkn$ at certain stationary points.
In the following, we will show that this conclusion  actually holds for all stationary points.

Let $\overline{f}(\Ponemn,\Ptwomn,\cdots,\PKn)$ denote $\sum_{k=1}^K w_k\olcalRkn$ the deterministic equivalent of $f$.
According to \eqref{eq:Gammakm_zero_mean_case} and \eqref{eq:tdetakmpri_zero_mean},
$\Gammakn$ can be written as
    \begin{IEEEeqnarray}{Cl}
       \Gammakn = \VMt\Sigmakn^2\VMt^H
    \end{IEEEeqnarray}
where $\Sigmakn^2$ is a diagonal matrix whose value depends on $\Ponemn,\Ptwomn, \cdots, \PKn$.
Then, we obtain the following theorem.
\vspace{-0.5em}
\begin{theorem}
\label{th:special_case_bdma}
Assume $\hatHkn=\mathbf{0}$. Then,
the left singular vector matrix of the linear precoders at the stationary points of the optimization problem
    \begin{IEEEeqnarray}{Cl}
        &\max_{\Ponemn,\cdots,\PKn}\overline{f}(\Ponemn,\Ptwomn,\cdots,\PKn)
        \nonumber \\
        &~~~~~~{\rm s.t.} ~~  \sumK \tr{\Pkn\Pkn^H} \leq P
        \label{eq:objective_optimization_problem_deterministic_equivalent}
    \end{IEEEeqnarray}
can be written as
    \begin{IEEEeqnarray}{Cl}
        \mathbf{U}_{\Pkn} = \VMt\Pikn
        \label{eq:bdma_special_case_optimal_linear_precoder}
    \end{IEEEeqnarray}
where $\Pikn$ is a permutation matrix.
\end{theorem}
\vspace{-0.5em}
\begin{IEEEproof}
The proof is provided in Appendix  \ref{sec:proof_of_special_case_bdma}.
\end{IEEEproof}
Theorem $\ref{th:special_case_bdma}$ proves the optimality of the beam domain transmission when $\hatHkn=\mathbf{0}$ and the objective function of the optimization problem \eqref{eq:objective_optimization_problem} is replaced by its deterministic equivalent.
Although the optimality of the beam domain transmission for single user MIMO has been well established in the literature \cite{veeravalli2005correlated, tulino2006capacity}, the optimality of the beam domain transmission for
multi-user MIMO still needs further investigation. In \cite{sunbeam}, the optimality of beam domain transmission for massive MIMO is proved under optimization an upper bound of the sum-rate.
In this paper, we prove the optimality of the beam domain transmission when the weighted sum-rate is replaced by its deterministic equivalent.

Using Theorem $\ref{th:special_case_bdma}$, we obtain that the optimal precoders can be written as
    \begin{IEEEeqnarray}{Cl}
        \Pkn  = \VMt\Pikn\Jkn\Vkn^H
        \label{eq:bdma_special_case_optimal_linear_precoder_1}
    \end{IEEEeqnarray}
where $\Jkn$ are $M_t \times d_k$ matrices with nonzero elements on the main diagonal and zeros elsewhere, and $\Vkn^H$ are any $d_k \times d_k$ unitary matrices.
Since $\Vkn^H$ has no impact on the expected weighted sum-rate, it can be set to a fixed unitary matrix.
For brevity, we set $\Vkn^H=\Idk$. Then, the optimal precoders can be rewritten as
    \begin{IEEEeqnarray}{Cl}
        \Pkn  = \VMt\Pikn\Jkn.
        \label{eq:bdma_special_case_optimal_linear_precoder_2}
    \end{IEEEeqnarray}
To achieve an algorithm with a complexity lower than Algorithm 2, we also set each $\Pikn$ to a fixed permutation matrix.
Let $\mathbf{a}_k$ be an $M_t \times 1$ row vector defined as
    \begin{IEEEeqnarray}{Cl}
        [\mathbf{a}_k]_j  = \sum\limits_{i=1}^{M_k}[\Omegak]_{ij} .
        \label{eq:bdma_special_case_optimal_linear_precoder_3}
    \end{IEEEeqnarray}
The permutation matrix $\Pikn$ is set to make the elements in $\mathbf{a}_k\Pikn$ are of descending order.
Then, we only need to optimize $\Jkn$ and the precoder design reduces to the power allocation optimization in the beam domain. Using the optimal structures of the precoders with fixed permutation matrices and the condition $\hatHkn=\mathbf{0}$, we obtain the following power allocation algorithm with lower complexity.
The inputs needed are $\Omegak$, $\Uk$ and $\sz$.
\hrule
\vspace{0.25em}
Algorithm 3: Linear precoder design when $\hatHkn=\mathbf{0}$
\vspace{0.25em}
\hrule
\begin{enumerate}[\IEEElabelindent=3em]
\item[Step 1:]
Set $d=0$. Initialize all the $\JknIterd$ with ones along the main diagonal and zeros elsewhere, and normalize them to satisfy the power constraint.
\item[Step 2:]
Calculate $\RknIterd$ according to
    \begin{IEEEeqnarray}{Cl}
        \RknIterd &=\sz\IMt  \nonumber \\
        &~+\sumnok \etaknpost{\VMt\Piln\JlnIterd(\JlnIterd)^H\Piln\VMt^H}. \nonumber
    \end{IEEEeqnarray}
\item[Step 3:]
Calculate $\Gammakn$ and $\tdGammakn$  according to \eqref{eq:Gammakm_zero_mean_case} and \eqref{eq:tdGammakm_zero_mean_case}.
\item[Step 4:]
Compute $\bdLambda_{\AknIterd}$, $\bdLambda_{\olCknIterd}$ and $\bdLambda_{\olFknIterd}$ according to
    \begin{IEEEeqnarray}{Cl}
    \bdLambda_{\AknIterd}&=\tdLambdakn{\invm{\RknIterd}} \nonumber \\
    \bdLambda_{\olCknIterd} &= \tdLambdakn{\invm{\RknIterd}} - \Sigmakn^2 \nonumber \\
    \bdLambda_{\olFknIterd} & = \Sigmakn^2 \nonumber \\
     &~- \invm{\IMt+\Sigmakn^2\Pikn\Jkn\Jkn^H\Pikn^H}\Sigmakn^2 \nonumber.
    \end{IEEEeqnarray}
\item[Step 5:] Update $ \Jkn^{(d+1)}$ by
    \begin{IEEEeqnarray}{Cl}
              \Jkn^{(d+1)} &= (\bdLambda_{\olDnIterd}+\mu^\star\IMt)^{-1} \nonumber \\
                            &~~~~    (w_k\bdLambda_{\AknIterd}+\bdLambda_{\olFknIterd})\JknIterd \nonumber
    \end{IEEEeqnarray}
where $\bdLambda_{\olDnIterd} =  \sum_{k=1}^K w_k\bdLambda_{\olCknIterd}$.
Set $d=d+1$.
\end{enumerate}
\hspace{1em}
Repeat Step 2 through Step 5 until convergence or until a
pre-set target is reached. Then the optimal precoders are obtained as $\Pkn^\star  = \VMt\Pikn\Jkn^\star$.
\vspace{0.5em}
\hrule
\vspace{0.5em}
In Algorithm 3, we have used the commutativity of the permutation matrices with the diagonal matrices to simplify the formulas.
Since $\bdLambda_{\olDnIterd}$ is a diagonal matrix, the $M_t \times M_t$ matrix inversion $(\bdLambda_{\olDnIterd}+\mu^\star\IMt)^{-1}$ in Algorithm 3 can be implemented
element-wisely. Thus, the computational complexity are further reduced when $\hatHkn=\mathbf{0}$.
In Algorithm 3, left multiplying $\VMt$ can be realized as several FFT operations. From \eqref{eq:eta_function_component} and \eqref{eq:etakmpri_zero_mean}, we obtain that the FFT operations $\VMt\Piln\JlnIterd$ in $\etaknpost{\VMt\Piln\JlnIterd(\JlnIterd)^H\Piln\VMt^H}$ do not need to be performed.
The only left multiplying $\VMt$ need to be performed is that in $\Pkn^\star  = \VMt\Pikn\Jkn^\star$, and its complexity is of order ${\cal{O}}(\frac{1}{2}KM_kM_t\log(M_t))$. For simplicity, we have assumed $M_t = 2^n$, where $n$ is an integer.
The computational complexity of the rest operations is of order ${\cal{O}}(K M_kM_t)$ per iteration.
When $K M_k$ is not small, the overall complexity of Algorithm 3 is not larger than that of RZF precoder which is of order
${\cal{O}}((K M_k)^2M_t+\frac{1}{2}(K M_k)^3)$.

\section{Simulation Results}
In this section, we provide simulation results to show the performance of the proposed algorithms.
We use the 3GPP stochastic channel model (SCM) \cite{WinnerScmImplementationIEEETranbst} to generate $\Uk$ and the channel power matrices $\Omegak$ as follows.
Denote  by $S$ the number of samples, and by ${\mathbf{H}}_{k}(s)$ the $s$-th sample of ${\mathbf{H}}_{k}$.
The sample average of  ${\mathbf{H}}_{k}$ is zero.
From the sample covariance matrix
${\mathbf{R}}_{r,{k}}=\frac{1}{S}\sum_{s=1}^S{\mathbf{H}}_{k}(s)({\mathbf{H}}_{k}(s))^H $
and its eigenvalue decomposition ${\mathbf{R}}_{r,{k}}={\mathbf{U}}_{k}{\mathbf{\Sigma}}_{r,{k}}{\mathbf{U}}_{k}^H$
the eigen-matrix $\mathbf{U}_{k}$ is obtained. Then, the coupling matrices $\boldsymbol{\Omega}_{k}$ is computed as \cite{weichselberger2006stochastic}
\begin{equation}
\boldsymbol{\Omega}_{k}=\frac{1}{S}\sum\limits_{s=1}^S\left(\mathbf{U}_{k}^H\mathbf{H}_{k}(s)\mathbf{V}_{M_t}\right)\odot\left({\mathbf{U}}_{k}^T{\mathbf{H}}_{k}(s)^*{\mathbf{V}}_{M_t}^*\right).
\end{equation}
The scenario used is ``urban\_marco". The
antenna arrays used at the BS and the UEs are both ULAs with $0.5\lambda$ spacing.
The shadow fading and path loss are not considered.  The
users in the cell are random uniformly distributed.
In all simulations, we set $P=1$, $w_k=1$, $d_k = M_k$ and $N_b=7$.  The $M_k$ for all the users are set to be the same. For simplicity, we set $\sigma_{\rm BS}^2=\sigma_z^2$.
The signal-to-noise ratio (SNR) is given by SNR$=\frac{1}{\sz}$.

\begin{figure}
\centering
\includegraphics[scale=0.58]{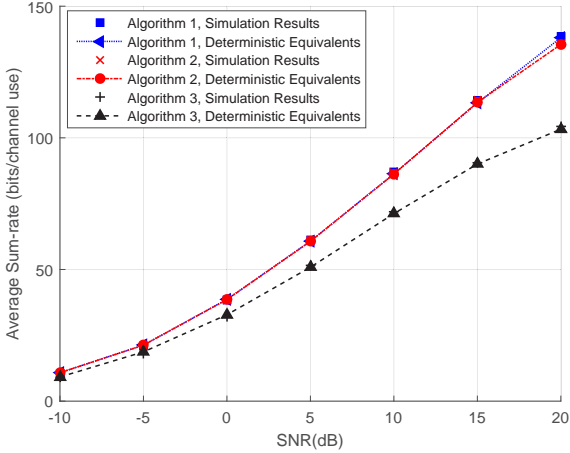}
\caption{Average sum-rate performance of the three proposed algorithms for a massive MIMO downlink with $M_t=64, M_k=4$, $K=10$, $\alpha_1 - \alpha_5=0.999$ and $\alpha_6 - \alpha_{10}=0.9$.}
\label{fig:Capacity_128Tx_10user_4Rx_Slow}
\end{figure}

We first investigate the performance of the three proposed algorithms. We consider a massive MIMO downlink system with $M_t=64, M_k=4$ and $K=10$. The values of $\alpha_1 - \alpha_5$ are set as $0.999$ and the values of $\alpha_6 - \alpha_{10}$ are set as $0.9$.
It indicates that the channels of the first 5 users are quasi-static, and that the other 5 users move slowly. Let $N_s$ denote the number of time slots used in the simulations.
Fig.~\ref{fig:Capacity_128Tx_10user_4Rx_Slow}
shows the simulation results of the average sum-rate performance of the three algorithms for this massive MIMO downlink over $N_s=100$ time slots. The number of iterations is set to $30$.
From Fig.~\ref{fig:Capacity_128Tx_10user_4Rx_Slow}, we see that the average sum-rates of the three algorithms increase almost linearly as the SNR increases. Furthermore, we see that the differences between the performance of Algorithms 1 and 2 are negligible. The performance gaps between Algorithm 3 and that of Algorithms 1 and 2 increase as the SNR increases. At SNR=$20$dB, the performance loss of Algorithm 3 is about $25$ percent. This is because Algorithm 3 is designed for the case when the channel means of all users equal zeros, and thus does not exploit the full benefits of the available statistical CSI at the BS.
We also present in Fig.~\ref{fig:Capacity_128Tx_10user_4Rx_Slow} the deterministic equivalents of the average sum-rate to show the accuracy of the deterministic equivalents. The deterministic equivalent results of all three algorithms are very accurate.

\begin{figure}
\centering
\includegraphics[scale=0.58]{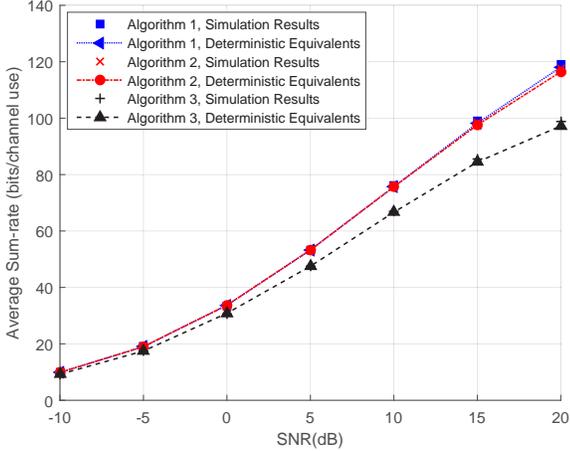}
\caption{Average sum-rate performance of the three proposed algorithms for a massive MIMO downlink with $M_t=64, M_k=4$, $K=10$ and the $\alpha$s presented in Table \ref{tb:alphas_for_scenario_2}.}
\label{fig:Capacity_128Tx_10user_4Rx_Fast}
\end{figure}

\begin{table}
\centering
\caption{The values of $\alpha_k$s in scenario 2.}
    \begin{tabular}{|c|c|c|c|c|}
      \hline
      $\alpha_1, \alpha_2$& $\alpha_3, \alpha_4$& $\alpha_5, \alpha_6$& $\alpha_7, \alpha_8$& $\alpha_9, \alpha_{10}$ \\
      \hline
      0.999 & 0.9 & 0.5 & 0.1 & 0 \\
      \hline
    \end{tabular}
\label{tb:alphas_for_scenario_2}
\end{table}

To investigate the performance of the three proposed algorithms when high speed users exist. We keep $M_t=64, M_k=4$, $K=10$ and $N_s=100$, but change the values of $\alpha_k$s to those presented in Table \ref{tb:alphas_for_scenario_2}.
Fig.~\ref{fig:Capacity_128Tx_10user_4Rx_Fast}
show the simulation results of the average sum-rate performance of the three algorithms for this scenario.
From Fig.~\ref{fig:Capacity_128Tx_10user_4Rx_Fast}, we see that the average sum-rates of the three algorithms are still increase almost linearly as the SNR increases, and that the differences between the performance of Algorithms 1 and 2 are also negligible. Furthermore, the performance gaps between Algorithm 3 and that of Algorithms 1 and 2 are become smaller. At SNR=$20$dB, the performance loss of Algorithm 3 is around $15$ percent.
This is because the difference between the statistical CSI that Algorithm 3 used and the available statistical CSI at the BS become smaller at this scenario. The deterministic equivalents of the average sum-rates are also presented in Fig.~\ref{fig:Capacity_128Tx_10user_4Rx_Fast} to show their accuracy. In this scenario, the deterministic equivalent results of all three Algorithms are also very accurate.

\begin{figure}
\centering
\includegraphics[scale=0.58]{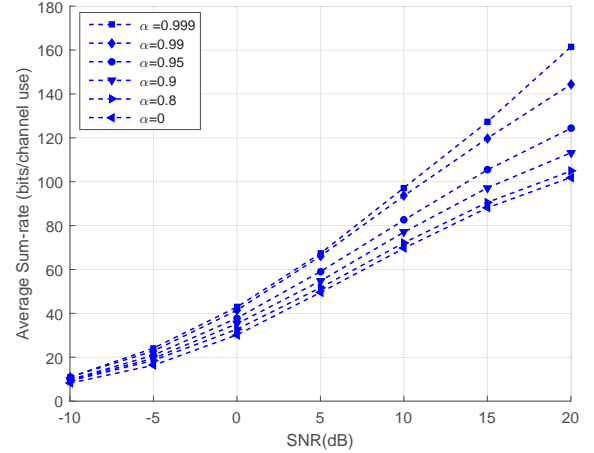}
\caption{Average sum-rate performance of Algorithm 1 for a massive MIMO downlink with $M_t=64, M_k=4$, $K=10$ for variant $\alpha$s.}
\label{fig:Capacity_128Tx_10user_4Rx_Variant}
\end{figure}

\begin{figure}
\centering
\includegraphics[scale=0.58]{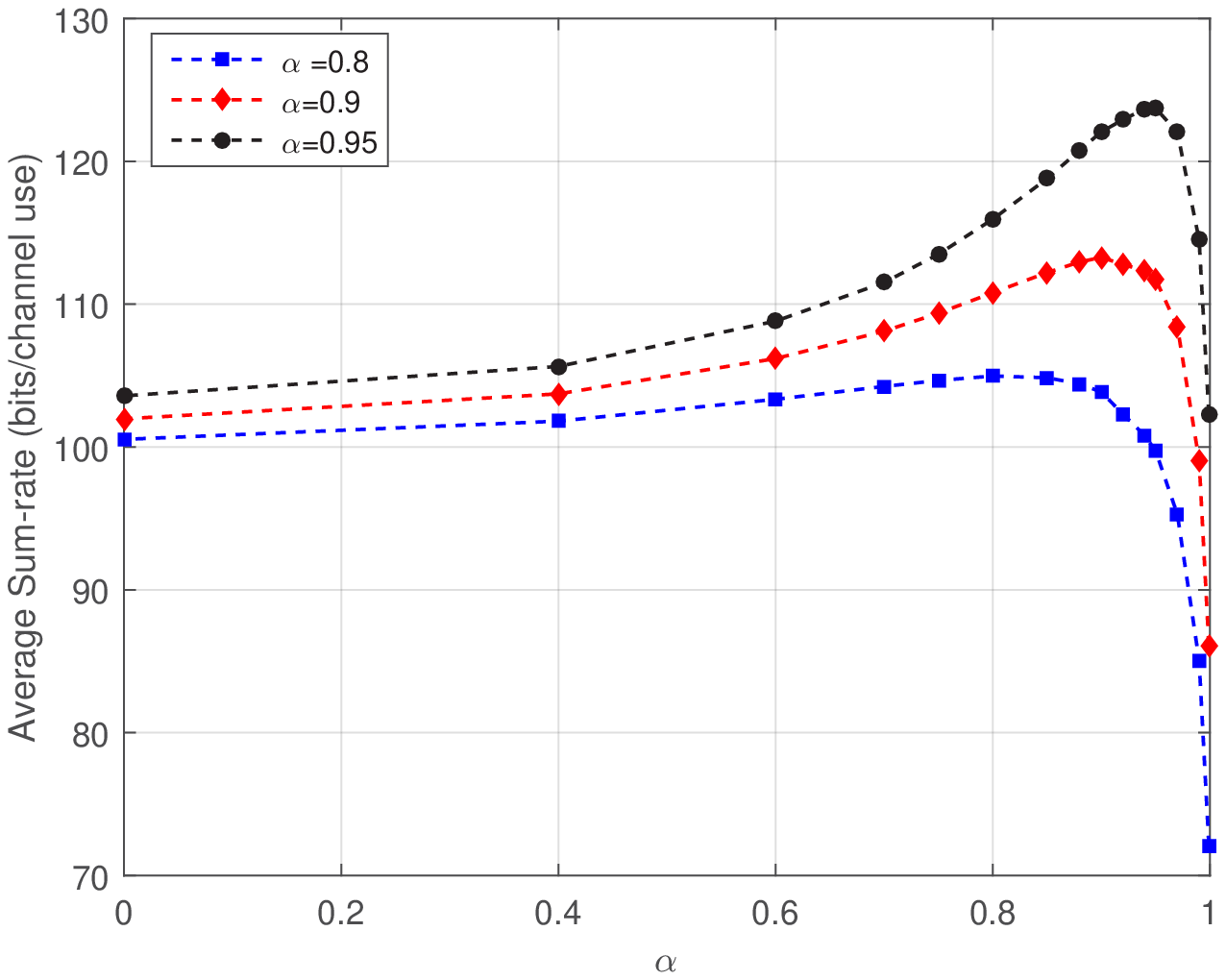}
\caption{Average sum-rate performance of Algorithm 1 for a massive MIMO downlink with $M_t=64, M_k=4$, $K=10$ at SNR$=20$dB for mismatched $\alpha$s.}
\label{fig:Capacity_128Tx_10user_4Rx_mismatch}
\end{figure}

By observing Fig.~\ref{fig:Capacity_128Tx_10user_4Rx_Slow} and Fig.~\ref{fig:Capacity_128Tx_10user_4Rx_Fast}, we realize that $\alpha_k$s affect the sum-rate performance. To show the influence of $\alpha_k$s on the
system performance, we set all the $\alpha_k$s to be the same $\alpha$ and simulate the average sum-rate performance of Algorithm 1. The $\alpha$s used in the simulation are $0.999, 0.99, 0.95, 0.9, 0.8$ and $0$. The simulated results for a massive MIMO downlink with $M_t=64, M_k=4$, $K=10$ are plotted in Fig.~\ref{fig:Capacity_128Tx_10user_4Rx_Variant}. From the simulation results, we observe that the sum-rate performance decreases as $\alpha$ becomes smaller.
This indicates that the system performance will degrade when the users move faster. Compared with that of $\alpha=0.999$, the sum-rate performance loss of $\alpha=0$ is about $40$ percent at SNR$=20$dB. Furthermore, the performance gap is relatively small between the sum-rate performance of $\alpha=0.8$ and $\alpha=0$. Thus, knowing the channel mean information can not bring much performance gain than only knowing the channel covariance information when $\alpha$ is smaller than $0.8$.
Since the value of $\alpha$ is important to the performance, we also show how sensitive the proposed precoder scheme is to the errors in the $\alpha$ in Fig.~\ref{fig:Capacity_128Tx_10user_4Rx_mismatch}. The true values of $\alpha$ are $0.95, 0.9$ and $0.8$. From Fig.~\ref{fig:Capacity_128Tx_10user_4Rx_mismatch}, we observe that the performance obtained when overestimate $\alpha$
is generally worse than that obtained by underestimate $\alpha$.

\begin{figure}
\centering
\includegraphics[scale=0.58]{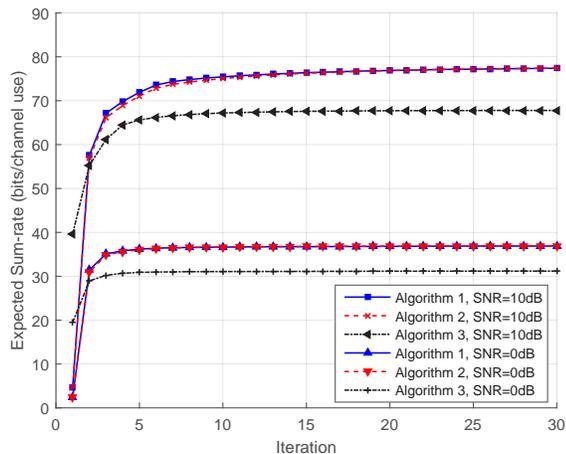}
\caption{Convergence trajectories of the three proposed algorithms for a massive MIMO downlink with $M_t=64, M_k=4$, $K=10$ and the $\alpha$s presented in Table \ref{tb:alphas_for_scenario_2}.}
\label{fig:Convergence_128Tx_10user_4Rx_Block_One}
\end{figure}

\begin{figure}
\centering
\includegraphics[scale=0.58]{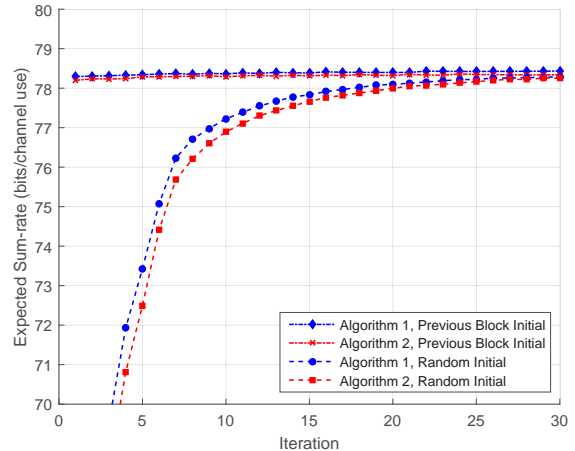}
\caption{Convergence trajectories of Algorithms 1 and 2 using different initials at the third block of the first time slot for a massive MIMO downlink at SNR$=10$dB with $M_t=64, M_k=4$, $K=10$ and the $\alpha$s presented in Table \ref{tb:alphas_for_scenario_2}.}
\label{fig:Convergence_128Tx_10user_4Rx_Block_Two}
\end{figure}
We then study the convergence behavior of the three proposed algorithms. The considered massive MIMO downlink is still that with $M_t=64, M_k=4$ and $K=10$, and the values of $\alpha_k$s are those presented in Table \ref{tb:alphas_for_scenario_2}. As shown in Step 1 in each algorithm,
we use random initializations for Algorithms 1 and 2, whereas the initializations for Algorithm 3 are fixed.
Fig.~\ref{fig:Convergence_128Tx_10user_4Rx_Block_One} shows the convergence behaviors of the three proposed algorithms at the second block of the first time slot for the massive MIMO downlink at two different SNRs. The expected sum-rate results presented in Fig.~\ref{fig:Convergence_128Tx_10user_4Rx_Block_One} are the deterministic equivalent results.
From Fig.~\ref{fig:Convergence_128Tx_10user_4Rx_Block_One}, we see that all three algorithms
quickly converge at SNR$=0$ dB and SNR$=10$dB. We also observe that
all three algorithms take more iterations to converge as the SNR increases. At SNR$=0$ dB, only $5$ iterations are need to obtain a good performance, whereas $15$ iterations are needed at SNR$=10$ dB.
Algorithm 3 only need to be performed once when the \textit{a priori} statistical CSI changes. Thus, the number of the iterations needed to make Algorithm 3 converge is not a problem. On the contrary, Algorithms 1 and 2 are performed once for each block. Thus, the number of the iterations needed to make Algorithms 1 and 2 converge is an issue.
To reduce the number of the iterations used in Algorithms 1 and 2,  we can use the resulting precoders from the previous block as initials instead of random initials at each block (not the first data block).
Fig.~\ref{fig:Convergence_128Tx_10user_4Rx_Block_Two} plots the convergence behaviors of Algorithms 1 and 2 at the third block of the first time slot using two different initials for the same massive MIMO downlink as that of Fig.~\ref{fig:Convergence_128Tx_10user_4Rx_Block_One} at SNR$=10$dB. From Fig.~\ref{fig:Convergence_128Tx_10user_4Rx_Block_Two}, we see that the resulting precoders from the previous block are very good initials, and thus only a few iterations are needed to achieve good performance and the computational complexity can be further reduced.

\begin{figure}
\centering
\includegraphics[scale=0.58]{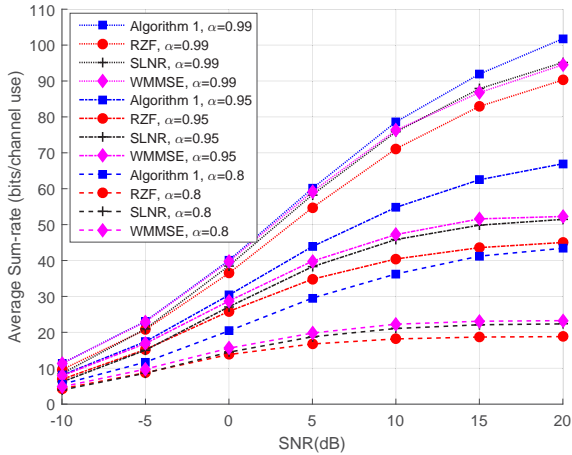}
\caption{Average sum-rate performance of Algorithm 1, the RZF precoder, the SLNR precoder and the iterative WMMSE precoder for a massive MIMO downlink with $M_t=64, M_k=1$ and $K=20$.}
\label{fig:Capacity_128Tx_20user_1Rx_Comparsion_with_others}
\end{figure}

Then, we investigate the performance of Algorithm 1 for massive MIMO downlinks with single antenna users.
We compare Algorithm 1 with the RZF, the SLNR and the iterative WMMSE precoders.
For the precoders except Algorithm 1, we use the perfect CSI from block one for all blocks at each slot.
We consider a massive MIMO downlink with $M_t=64$, $M_k=1$ and $K=20$. The $\alpha_k$s used in the simulations are the same $\alpha$ with $\alpha=0.99, 0.95$ and $0.8$.
Fig.~\ref{fig:Capacity_128Tx_20user_1Rx_Comparsion_with_others} plots the average sum-rate performance of four algorithms over $N_s=100$ time slots.
From the simulation results, we observe that
Algorithm 1 can achieve much better performance than those of other precoderes at all three cases.
Furthermore, the performance gain becomes more significant as the CSI becomes more inaccurate.
Thus, Algorithm 1 is more effective in improving the sum-rate performance for massive MIMO with imperfect CSI than the other precoders.

\begin{figure}
\centering
\includegraphics[scale=0.58]{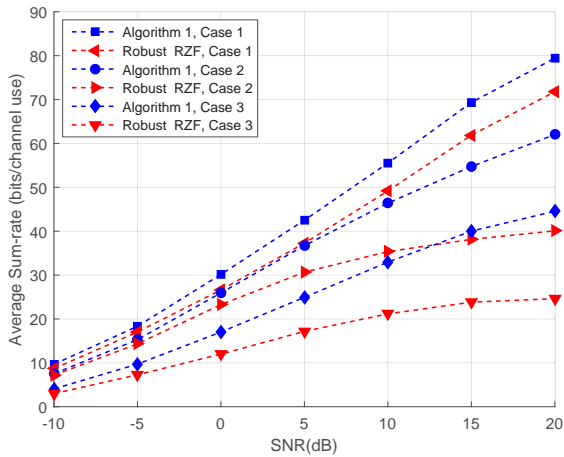}
\caption{Average sum-rate performance of Algorithm 1 and the robust RZF precoder for a massive MIMO downlink with $M_t=64, M_k=1$ and $K=10$.}
\label{fig:Capacity_128Tx_20user_1Rx_Comparsion_with_RZF}
\end{figure}

\begin{table}
\centering
\caption{The values of $\alpha_k$s in scenario 3.}
    \begin{tabular}{|c|c|c|c|c|c|c|}
      \hline
      & $\alpha_1, \alpha_2$& $\alpha_3, \alpha_4$& $\alpha_5$¡¡&$\alpha_{6}$& $\alpha_{7},\alpha_{8}$& $\alpha_{9}, \alpha_{10}$ \\
      \hline
      Case 1 & 0.999 & 0.999 & 0.999 & 0.999 & 0.999 & 0.999 \\
      \hline
      Case 2 & 0.999 & 0.999 & 0.999 & 0.9  & 0.9 & 0.9 \\
      \hline
      Case 3 & 0.999 & 0.9 & 0.5 & 0.5 & 0.1 & 0 \\
      \hline
    \end{tabular}
\label{tb:alphas_for_scenario_3}
\end{table}

Finally, we compare the sum-rate performance of Algorithm 1 with that of the robust RZF precoder.
We consider a massive MIMO downlink with $M_t=64$ transmit antennas at the BS and $K=10$ single antenna users. The $\alpha_k$s used in the simulations are presented in Table \ref{tb:alphas_for_scenario_3}.
Fig.~\ref{fig:Capacity_128Tx_20user_1Rx_Comparsion_with_RZF} plots the average sum-rate performance of Algorithm 1 and the robust RZF for the considered scenario over $N_s=100$ time slots.
As shown in Fig.~\ref{fig:Capacity_128Tx_20user_1Rx_Comparsion_with_RZF},
Algorithm 1 can achieve much better performance than that of the robust RZF precoder at all three cases.
Furthermore, we observe that the performance gains are small at
low SNR, but become significant as SNR increases. The reason behind the performance gain of the proposed precoder is as follows.
While the robust RZF are designed by minimizing and maximizing the average MSE,
the proposed precoder of one user is obtain iteratively by considering how the precoder will affect the rate performance of other users when their precoders are known.
To put it simply, the proposed design iteratively adjust the interference to each other to improve the sum-rate performance directly.

\section{Conclusion}
In this paper, we investigated the design of robust linear precoders for the massive MIMO downlink with imperfect CSI. The available imperfect CSI for each UE obtained at the BS is modeled
as statistical CSI under a jointly correlated channel model with both channel mean and channel variance information, which includes the effects of channel estimation error, channel aging and spatial correlation. We derived an algorithm for the linear precoder design by using the MM algorithm. The derived algorithm can achieve a stationary point of the expected weighted sum-rate maximization problem. We then used the deterministic equivalent method to compute the approximations of several key matrices used in the robust linear precoder design. Then, we proposed an
algorithm for robust linear precoder design based on the deterministic equivalent method.
The proposed algorithm needs $2K$ large dimensional matrix inversions per iteration. To reduce the computational complexity, we then derived two low-complexity algorithms, one for the general case, and the other for the case when all the channel means are zeros. For the late case, we also proved the optimality of the beam domain transmissions and  the precoder design reduced to power allocation optimization in the beam domain. Simulation results showed that the proposed robust linear precoder designs apply to various mobile scenarios and achieve high spectral efficiency.

We conclude this paper by providing two possible topics for future research of this work.
The channel model of this paper is established by assuming that the ULA antenna is used.
In practical massive MIMO systems, uniform planar array (UPA) antennas are also widely used.
Thus, an extension of this work would be to consider the massive MIMO with a UPA antenna.
In this paper, we focused on the precoder design for TDD massive MIMO systems.
Another extension of this work would be to consider also the impact of limited feedback, which is essential in FDD massive MIMO systems.
Consider an example when the channel feedback is obtained from quantization on the Grassmann manifold.
From the finite feedback of the channels, we can obtain a probability distribution of the channels on the Grassmann manifold.
Using statistics on Riemannian manifold \cite{pennec2006intrinsic} to replace the statistics computed from the established channel model in the proposed methods, it might be able to obtain a precoder design which is robust to quantization error induced by limited feedback.

\appendices

\section{Proof of Theorem \ref{th:minorizing_function_1}}
\label{sec:proof_of_minorizing_function_1}
Let the matrix $\Ekn$ be defined as $\Ekn=\invm{\Idk + \Pkn^H\Hkn^H\Rkn^{-1} \Hkn\Pkn}$.
Then, the ergodic rate of the $k$-th user can be rewritten as $\calRkn=\mean{\logdet{\Ekn^{-1}}}$.
The function $\mean{\logdet{\Ekn^{-1}}}$ is a convex function of $\Ekn$ on $\mathbf{S}^n_{++}$.
Let $\EknIterd$  be defined as
    \begin{IEEEeqnarray}{Cl}
        \EknIterd&=\invm{\Idk + (\PknIterd)^H\Hkn^H(\RknIterd)^{-1} \Hkn\PknIterd}.
        \label{eq:definition_of_EkmIterd}
    \end{IEEEeqnarray}
Using the first order condition of convex functions,
we obtain
    \begin{IEEEeqnarray}{Cl}
        \mean{\logdet{\Ekn^{-1}}}
        &\geq \mean{\logdet{\invm{\EknIterd}}} \nonumber \\
        &~~- \mean{\tr{\invm{\EknIterd}(\Ekn-\EknIterd)}}
        \nonumber \\
        &= c_{kn}'  - \mean{\tr{\invm{\EknIterd}\Ekn}}
        \label{eq:mutual_information_inequality_1}
    \end{IEEEeqnarray}
where $c_{kn}'=\mean{\logdet{\invm{\EknIterd}}} + \tr{\Idk}$ is a constant.
The step in equation \eqref{eq:mutual_information_inequality_1} is inspired by \cite{razaviyayn2013unified}.
The item $-\mean{\tr{(\EknIterd)^{-1}\Ekn}}$ is still not a simple function of the precoding matrices.
Inspired by \cite{shi2011iteratively}, let $\Gkn^H$ denote the linear receiver of the $k$-th user.
The mean-square error (MSE) matrix of the $k$-th estimate
$\hatxkn=\Gkn^H\ykn$ is given by
    \begin{IEEEeqnarray}{Cl}
          \!\!\!\!\!\!\!\!\Thetakn &= \mean{(\hatxkn - \xkn)(\hatxkn - \xkn)^H}
                \nonumber\\
                &= (\Idk - \Gkn^H\Hkn\Pkn)(\Idk - \Gkn^H\Hkn\Pkn)^H
                \nonumber \\
                &+ \Gkn^H\sumnok \mean{\Hkn\Pln\Pln^H\Hkn^H}\Gkn +  \sz\Gkn^H\Gkn.
          \label{eq:MSE_matrix_M}
    \end{IEEEeqnarray}
From \eqref{eq:MSE_matrix_M}, we observe that the function $\tr{(\EknIterd)^{-1}\Thetakn}$ is a convex function of $\Gkn$ and its global minimum is achieved when $(\Gkn^\star)^H$
is the linear minimum mean-square error (MMSE) receiver, \textit{i.e.},
$(\Gkn^\star)^H =\Pkn^H\Hkn^H\invm{\Rkn + \Hkn\Pkn\Pkn^H\Hkn^H}$.
Substituting $\Gkn^\star$ into \eqref{eq:MSE_matrix_M}, we obtain $\Thetakn|_{\Thetakn=\Gkn^\star}=\Ekn$.
Thus, we have
    \begin{IEEEeqnarray}{Cl}
          \mean{\tr{\invm{\EknIterd}\Ekn}} \leq \mean{\tr{\invm{\EknIterd}\Thetakn}}
          \label{eq:minimum_mse}
    \end{IEEEeqnarray}
for any $\Gkn$.
From \eqref{eq:mutual_information_inequality_1} and \eqref{eq:minimum_mse}, we obtain
    \begin{IEEEeqnarray}{Cl}
        \mean{\logdet{\Ekn^{-1}}}
        &\geq c_{kn}' - \mean{\tr{\invm{\EknIterd}\Thetakn}}.
        \label{eq:mutual_information_inequality_2}
    \end{IEEEeqnarray}
To make the equality in  \eqref{eq:mutual_information_inequality_2} hold at $\PonemnIterd, \cdots, \PKnIterd$, we set $\Gkn=\GknnIterd$, which is defined by
    \begin{IEEEeqnarray}{Cl}
        &(\GknnIterd)^H \nonumber \\
        &~~=(\PknIterd)^H\Hkn^H\invm{\RknIterd + \Hkn\PknIterd(\PknIterd)^H\Hkn^H}.
                \label{eq:check_Uk_at_check_Pk}
    \end{IEEEeqnarray}
When $\Gkn = \GknnIterd$,
we obtain from \eqref{eq:MSE_matrix_M} that
     \begin{IEEEeqnarray}{Cl}
         &- \mean{\tr{\invm{\EknIterd}\Thetakn}} \nonumber \\
         &~=- c_{kn}'' +   \tr{(\AknIterd\PknIterd)^H\Pkn} +  \tr{\AknIterd\PknIterd\Pkn^H}
        \nonumber \\
        &~~~-  \tr{\BknIterd\Pkn\Pkn^H+\CknIterd\sumnok\Pln\Pln^H}.
        \label{eq:expansion_of_Ekm_inverse_times_Thetakm}
    \end{IEEEeqnarray}
where $c_{kn}''$, $\AknIterd\PknIterd$, $\BknIterd$ and $\CknIterd$ are defined as
\begin{IEEEeqnarray}{Cl}
    c_{kn}''&=\mean{\tr{\invm{\EknIterd}}} \nonumber \\
    &~~- \sz\mean{\tr{\invm{\EknIterd}(\GknnIterd)^H\GknnIterd}}
\\
        \!\!\!\!\!\!\!\!(\AknIterd\PknIterd)^H&=\mean{\invm{\EknIterd}(\GknnIterd)^H\Hkn}
        \label{eq:Akm_at_iterd}
        \\
        \BknIterd&=\mean{\Hkn^H\GknnIterd\invm{\EknIterd}(\GknnIterd)^H\Hkn}
        \label{eq:Bkm_at_iterd}
        \\
        \CknIterd&=\mean{\Hkn^H{\mean{\GknnIterd\invm{\EknIterd}(\GknnIterd)^H}}\Hkn}.
        \label{eq:Ckm_at_iterd}
    \end{IEEEeqnarray}
Furthermore, we define
\begin{IEEEeqnarray}{Cl}
        \cknIterd&=c_{kn}'
                - c_{kn}''.
        \label{eq:ckm_at_iterd}
    \end{IEEEeqnarray}
From \eqref{eq:mutual_information_inequality_2},
\eqref{eq:expansion_of_Ekm_inverse_times_Thetakm} and
\eqref{eq:ckm_at_iterd} to \eqref{eq:Ckm_at_iterd}, we obtain
\begin{IEEEeqnarray}{Cl}
        &\mean{\logdet{\Ekn^{-1}}} \nonumber \\
        &~\geq  \cknIterd +   \tr{\AknIterd\Pkn(\PknIterd)^H} +  \tr{\AknIterd\PknIterd\Pkn^H}
        \nonumber \\
        &~~~~~~~~-  \tr{\BknIterd\Pkn\Pkn^H+\CknIterd\sumnok\Pln\Pln^H}.
        \label{eq:rate_k_condition_1}
    \end{IEEEeqnarray}
where the equality can be achieve at the fixed precoding matrices $\PonemnIterd,\PtwomnIterd, \cdots, \PKnIterd$.

Substituting \eqref{eq:definition_of_EkmIterd} and \eqref{eq:check_Uk_at_check_Pk} into \eqref{eq:Akm_at_iterd}, we obtain
    \begin{IEEEeqnarray}{Cl}
        \!\!\!\!\!\!\AknIterd&=\mean{(\Hkn^H\invm{\ckRknIterd}\Hkn}
           \nonumber \\
           &~+\mean{\Hkn^H\invm{\RknIterd}(\ckRknIterd-\RknIterd)
        \invm{\ckRknIterd}\Hkn}.
        \label{eq:expansion_of_Ak1}
    \end{IEEEeqnarray}
From \eqref{eq:expansion_of_Ak1}, we then obtain the expression of $\AknIterd$ in \eqref{eq:theorem_AkmIterd}.
Similarly, we obtain the expression of $\BknIterd$ in \eqref{eq:theorem_BkmIterd} and the expression of  $\CknIterd$ in \eqref{eq:theorem_CkmIterd}.
Let $\DknIterd$ be defined as $\DknIterd =  w_k\BknIterd + \sum_{l\ne k}^K w_l\ClnIterd$.
Recall that
$f(\Ponemn,\Ptwomn, \cdots, \PKn)
                =\sum_{k=1}^K w_k\mean{\logdet{(\Ekn)^{-1}}}$.
From \eqref{eq:rate_k_condition_1} we obtain the
function $g_1$ defined in \eqref{eq:minorizing_function_definition_1} is a minorizing function of the objective function.

\section{Proof of Theorem \ref{th:deterministic_equivalent_of_mean_Bk_and_Ck} }
\label{sec:proof_of_deterministic_equivalent_of_mean_Bk_and_Ck}
Using methods similar to that in the proof of Theorem 4  in \cite{alu2016free}, we obtain
\begin{IEEEeqnarray}{Cl}
        &\left.\frac{\partial \olcalRkn}{\partial (\Pkn\Pkn^H)}\right|_{\Pkn=\PknIterd} \nonumber
        \\
        &~= (\IMt+\Gammakn\PknIterd(\PknIterd)^H)^{-1}\Gammakn.
        \label{eq_partial_olcalRkm_wst_PkmPkmH}
\end{IEEEeqnarray}
From \eqref{mutual_information}, we obtain the gradient of $\calRkn$ with respect to $\Pkn\Pkn^H$ as
\begin{IEEEeqnarray}{Cl}
&\left.\frac{\partial \calRkn}{\partial (\Pkn\Pkn^H)}\right|_{\Pkn=\PknIterd}
=\meanc{}{\Hkn^H\invm{\ckRknIterd}\Hkn}.
\label{eq_partial_calRkm_wst_PkmPkmH}
\end{IEEEeqnarray}
Since $\olcalRkn$ is the deterministic equivalent of $\calRkn$, we obtain from
\eqref{eq_partial_olcalRkm_wst_PkmPkmH} and \eqref{eq_partial_calRkm_wst_PkmPkmH} that the matrix $\olBknIterd$ provided in
\eqref{eq:BkmIterd_deterministic_equivalent} is  the deterministic equivalent
of $\BknIterd$.

Similarly, we can obtain the gradient of $\olcalRkn$ with respect to $\Rkn$ from \eqref{eq:deterministic_equivalent_of_user_k_rate_2}  as
\begin{IEEEeqnarray}{Cl}
        \frac{\partial \olcalRkn}{\partial \Rkn} &= -\Rkn^{-1}(\IMk+\tdGammakn\Rkn^{-1})^{-1}\tdGammakn\Rkn^{-1}.
\end{IEEEeqnarray}
The gradients of $\olcalRkn$ with respect to $\Pln\Pln^H$, $l \ne k$, are then obtained from the above equation.
Using a method similar to that in Lemma 4 of \cite{xiao2011globally}, we then obtain
\begin{IEEEeqnarray}{Cl}
       &\left.\frac{\partial \olcalRkn}{\partial (\Pln\Pln^H)}\right|_{\Pln=\PlnIterd}
       \nonumber \\
       &~=\meanc{}{\Hkn^H((\RknIterd+\tdGammakn)^{-1}-\invm{\RknIterd})\Hkn}
       \nonumber \\
       &~ =-\olCknIterd.
       \label{eq_partial_olcalRkm_wst_PlmPlmH}
\end{IEEEeqnarray}
From \eqref{mutual_information} and the chain rule, we then obtain
    \begin{IEEEeqnarray}{Cl}
         &\left.\frac{\partial \calRkn}{\partial (\Pln\Pln^H)}\right|_{\Pln=\PlnIterd}
        \nonumber \\
        &~ =\meanc{}{\Hkn^H(\meanc{}{\invm{\ckRknIterd}}-\invm{\RknIterd})\Hkn}
        \nonumber \\
        &~ =-\CknIterd.
        \label{eq_partial_calRkm_wst_PlmPlmH}
    \end{IEEEeqnarray}
From \eqref{eq_partial_olcalRkm_wst_PlmPlmH}  and \eqref{eq_partial_calRkm_wst_PlmPlmH}, we obtain $\olCknIterd$
is the deterministic equivalent
of $\CknIterd$. Thus, \eqref{eq:CkmIterd_deterministic_equivalent} holds.

\section{Proof of Theorem \ref{th:minorizing_function_2}}
\label{sec:proof_of_minorizing_function_2}
We first rewrite the minorizing function $g_1$ provided by Theorem \ref{th:minorizing_function_1} as
\begin{IEEEeqnarray}{Cl}
        g_1&= \sumK w_k\cknIterd +  \sumK w_k\tr{\AknIterd\Pkn(\PknIterd)^H} \nonumber \\
        & +\sumK  w_k\tr{\AknIterd\PknIterd\Pkn^H}
        \nonumber \\
        &~~- \sumK \tr{(\DknIterd+\FknIterd)\Pkn\Pkn^H}
        \nonumber \\
        &~~~+ \sumK \tr{\FknIterd\Pkn\Pkn^H}
        \label{eq:minorizing_function_definition_1_variation_1}.
    \end{IEEEeqnarray}
The fourth item on the RHS of the equality of \eqref{eq:minorizing_function_definition_1_variation_1} is a convex quadratic function of the precoding matrices. Using the first order condition of convex functions,
we obtain
\begin{IEEEeqnarray}{Cl}
        &\sumK \tr{\FknIterd\Pkn\Pkn^H} \geq \sumK \tr{\FknIterd\PknIterd(\PknIterd)^H}
        \nonumber \\
        &~~+  \sumK \tr{\FknIterd(\Pkn-\PknIterd)(\PknIterd)^H}
        \nonumber \\
        &~~~~ +\sumK \tr{\FknIterd\PknIterd(\Pkn-\PknIterd)^H}.
            \label{eq:minorizing_function_definition_1_variation_2}
    \end{IEEEeqnarray}
From \eqref{eq:minorizing_function_definition_1_variation_1} and \eqref{eq:minorizing_function_definition_1_variation_2},
we then obtain
\begin{IEEEeqnarray}{Cl}
        g_1 &\geq \cnIterd
        +  \sumK \tr{(w_k\AknIterd+\FknIterd)\Pkn(\PknIterd)^H}
        \nonumber \\
        &~~     +\sumK  \tr{(w_k\AknIterd+\FknIterd)\PknIterd\Pkn^H}
        \nonumber \\
        &~~~~- \sumK \tr{(\DknIterd+\FknIterd)\Pkn\Pkn^H}
        \label{eq:minorizing_function_definition_1_variation_3}.
    \end{IEEEeqnarray}
where $\cnIterd$ is defined as
    \begin{IEEEeqnarray}{Cl}
        \cnIterd = \sumK w_k\cknIterd - \sumK \tr{\FknIterd\PknIterd(\PknIterd)^H}.
        \label{eq:constant_cn}
    \end{IEEEeqnarray}
Let $g_2$ be
defined as in \eqref{eq:minorizing_function_definition_1_variation_1}. From
\eqref{eq:minorizing_function_definition_1_variation_3}, we have
    \begin{IEEEeqnarray}{Cl}
        &g_2(\Ponemn,\Ptwomn, \cdots, \PKn)
         \leq g_1(\Ponemn,\Ptwomn, \cdots, \PKn).
    \end{IEEEeqnarray}
Furthermore, it is easy to verify that the equality is achieved at $\PonemnIterd,\PtwomnIterd, \cdots, \PKnIterd$.
Thus, $g_2$ is also a minorizing function of the objective function.

\vspace{-1em}
\section{Proof of Theorem \ref{th:special_case_bdma}}
\label{sec:proof_of_special_case_bdma}
Since $\overline{f}$ denotes $\sum_{k=1}^K w_k\olcalRkn$, we obtain
\begin{IEEEeqnarray}{Cl}
       \frac{\partial \overline{f}}{\partial \Pkn^*} &=
       (\IMt+\Gammakn\Pkn\Pkn^H)^{-1}\Gammakn\Pkn \nonumber \\
       &~~-\sumnok\tdetalnpost{ \Rln^{-1}- (\Rln+\tdGammaln)^{-1}}\Pkn.
\end{IEEEeqnarray}
We define the Lagrangian as
    \begin{IEEEeqnarray}{Cl}
        &\mathcal{L}(\mu, \Ponemn,\Ptwomn, \cdots, \PKn)
        \nonumber \\
        &~= - \overline{f} + \mu(\sumK \tr{\Pkn\Pkn^H} - P) .
        \label{eq:lagrangian_of_deterministic_problem_1}
    \end{IEEEeqnarray}
From the first order optimal conditions of \eqref{eq:lagrangian_of_deterministic_problem_1}, we obtain
\begin{IEEEeqnarray}{Cl}
        & -w_k(\IMt+\Gammakn\Pkn\Pkn^H)^{-1}\Gammakn\Pkn \nonumber \\
        & ~~ + \sumnok w_l\tdetalnpost{ \Rln^{-1}- (\Rln+\tdGammaln)^{-1}}\Pkn \nonumber \\
        &~~~~+ \mu\Pkn = \mathbf{0}.
       \label{eq:first_order_condition_of_lagrangian_of_deterministic_problem_1}
\end{IEEEeqnarray}
From \eqref{eq:tdetakmpri_zero_mean}, we obtain
\begin{IEEEeqnarray}{Cl}
       &\!\!\!\!\sumnok w_l\tdetalnpost{ \Rln^{-1}- (\Rln+\tdGammaln)^{-1}}
       =\VMt\tdSigmakn^2\VMt^H
\end{IEEEeqnarray}
where $\tdSigmakn^2$ is a diagonal matrix.
Then, the first order conditions in \eqref{eq:first_order_condition_of_lagrangian_of_deterministic_problem_1} become
\begin{IEEEeqnarray}{Cl}
       &w_k(\IMt+\Gammakn\Pkn\Pkn^H)^{-1}\Gammakn\Pkn \nonumber \\
       &~~= \VMt\tdSigmakn^2\VMt^H\Pkn + \mu\Pkn.
       \label{eq:first_order_condition_of_lagrangian_of_deterministic_problem_1_variation_1}
\end{IEEEeqnarray}
When $\hatHkn=\mathbf{0}$, we have $\Gammakn=\VMt\Sigmakn^2\VMt^H$.
We define $\Tkn=\mu\IMt+\VMt\tdSigmakn^2\VMt^H$, $\Gammakn'=\Tkn^{-1/2}\Gammakn\Tkn^{-1/2}$ and $\Pkn'=\Tkn^{1/2}\Pkn$.
Then, the conditions in \eqref{eq:first_order_condition_of_lagrangian_of_deterministic_problem_1_variation_1} become
\begin{IEEEeqnarray}{Cl}
       w_k\Gammakn'\Pkn'(\IMt+(\Pkn')^H\Gammakn'\Pkn')^{-1} = \Pkn'.
       \label{eq:first_order_condition_of_lagrangian_of_deterministic_problem_1_variation_2}
\end{IEEEeqnarray}
Right multiplying both sides of \eqref{eq:first_order_condition_of_lagrangian_of_deterministic_problem_1_variation_2} by the item $(\IMt+(\Pkn')^H\Gammakn'\Pkn')(\Pkn')^H$, we obtain
\begin{IEEEeqnarray}{Cl}
       &w_k\Gammakn'\Pkn'(\Pkn')^H \nonumber \\
       &~~= \Pkn'(\IMt+(\Pkn')^H\Gammakn'\Pkn')(\Pkn')^H.
       \label{eq:first_order_condition_of_lagrangian_of_deterministic_problem_1_variation_4}
\end{IEEEeqnarray}
Thus, we obtain $\Gammakn'\Pkn'(\Pkn')^H=\Pkn'(\Pkn')^H\Gammakn'$, which indicates $\Pkn'(\Pkn')^H$ commutes
with $\Gammakn'$. From Theorem 9-33 of \cite{perlis1991theory} we then obtain $\Pkn'(\Pkn')^H$ and
$\Gammakn'$ have the same eigenvectors. From $\Gammakn'=\Tkn^{-1/2}\Gammakn\Tkn^{-1/2}$, we have that the eigenvectors
of $\Gammakn'$ and $\Gammakn$ are the same. Thus, the left singular vector matrix of  $\Pkn'$ can be written as $\mathbf{U}_{\Pkn'} = \VMt\Pikn'$,
where $\Pikn'$ is a permutation matrix. From $\Pkn=\Tkn^{-1/2}\Pkn'$ and $\mathbf{U}_{\Pkn'} = \VMt\Pikn'$, we obtain \eqref{eq:bdma_special_case_optimal_linear_precoder} holds finally.

\section*{Acknowledgment}
We would like to thank the editor and the anonymous
reviewers for their helpful comments and suggestions.

\bibliographystyle{IEEEtran}
\bibliography{IEEEabrv,this_reference}

\end{document}